\documentclass[aps,prd,amsmath,amssymb,preprintnumbers,preprint,nofootinbib,a4paper,11pt]{revtex4-1}
\pdfoutput=1
\usepackage[utf8]{inputenc} 
\usepackage{graphicx}
\graphicspath{{./figures/}}
\usepackage{url}
\usepackage[bookmarks, pagebackref=false]{hyperref}
\usepackage[usenames,dvipsnames]{xcolor}
\definecolor{orange}{cmyk}{0,0.5,1,0}
\definecolor{rossoCP3}{cmyk}{0,.88,.77,.40}
\definecolor{graa}{rgb}{0.8,0.8,0.8}
\definecolor{blaa}{rgb}{0.2,0.2,0.6}
		\hypersetup{
			colorlinks, 
			bookmarksopen, 
			bookmarksnumbered,
			citecolor=blaa, 		
			linkcolor=rossoCP3,	
			urlcolor=rossoCP3,			
			}
\usepackage{bm}
\usepackage{bbm}
\usepackage{pxfonts}

\usepackage{placeins}
\usepackage{xspace}
\usepackage{cancel} 

\usepackage{slashed}
\usepackage[caption=false, labelformat=simple, listofformat=subsimple, labelfont=default, margin=5pt, justification=raggedright]{subfig}

	\makeatletter
		\renewcommand{\p@subfigure}{}
	\makeatother

\usepackage{natbib}

\newcommand{\ea}[1]{
\begin{align}
#1
\end{align}
}

\newcommand{\beq}{\begin{eqnarray}}
\newcommand{\eeq}{\end{eqnarray}}

\newcommand{\p}{\partial}

\newcommand{\bmp}{\noindent\begin{minipage}{16cm}}
\newcommand{\emp}{\end{minipage}\vskip 7mm} 

\DeclareMathOperator{\Tr}{Tr}

\newcommand{\mc}{\mathcal}

\def\lsim{\mathrel{\rlap{\lower4pt\hbox{\hskip1pt$\sim$}}
    \raise1pt\hbox{$<$}}}                
\def\gsim{\mathrel{\rlap{\lower4pt\hbox{\hskip1pt$\sim$}}
    \raise1pt\hbox{$>$}}}                

\newcommand{\asBox}{\raisebox{-2.5pt}{$\Box$}\!\!\!\!\raisebox{2.5pt}{$\Box$}}
\newcommand{\sBox}{\Box\!\Box}

\begin{document}
\title{\Large  \color{rossoCP3}Asymptotically safe and free chiral theories with and without scalars} 
 
\author{Esben {\sc M\o lgaard}}
\email{molgaard@cp3.sdu.dk} 
\author{Francesco {\sc Sannino}}
\email{sannino@cp3.dias.sdu.dk}
 \affiliation{
{ \rm CP}$^{ \bf 3}${\rm-Origins} \& the Danish Institute for Advanced Study {\rm D-IAS},\\ 
\mbox{University of Southern Denmark, Campusvej 55, DK-5230 Odense M, Denmark}\\
}

\begin{abstract}
We unveil the dynamics of four dimensional chiral gauge-Yukawa theories featuring several scalar degrees of freedom transforming according to distinct representations of the underlying gauge group. We consider generalized Georgi-Glashow and Bars-Yankielowicz theories. We determine, to the maximum known order in perturbation theory, the phase diagram of these theories and further disentangle their ultraviolet asymptotic nature according to whether they are asymptotically free or safe.   
We therefore extend the number of theories that are known to be fundamental in the Wilsonian sense to the case of chiral gauge theories with scalars.  

\vspace{0.5cm}
\noindent
{ \footnotesize  \it Preprint: CP$^3$-Origins-2016-040 DNRF90}
\end{abstract}

\maketitle

\section{Chiral gauge-Yukawa theories}
The Standard Model of particle interactions is a chiral gauge-Yukawa field theory. These theories therefore play an important role in nature. In addition some of the first and most compelling attempts to unify the electromagnetic, weak and color interactions make use of chiral gauge-Yukawa theories with a single gauge coupling.

However, very little is known about the interacting dynamics of this kind of theories. Furthermore, their being chiral makes it impossible, at the moment, to investigate their dynamics via first principle lattice simulations. These are the reasons that compel us to uncover in this paper some of the key dynamical properties of these theories via higher order computations. Our theories contain besides chiral fermions also several kind of scalars transforming according to different representations of the underlying gauge and global symmetries.  We will concentrate on important ultraviolet and infrared properties of the theories such as, for example, whether the theories are completely asymptotically free \cite{Gross:1973ju,Cheng:1973nv,Callaway:1988ya,Holdom:2014hla,Giudice:2014tma,Pica:2016krb} or safe \cite{Weinberg:1980gg,Litim:2014uca,Litim:2015iea,Intriligator:2015xxa,Martin:2000cr}. In both scenarios, i.e. asymptotic freedom or safety\footnote{Asymptotic safety has also been invoked  \cite{Weinberg:1980gg} to help taming quantum gravity problems \cite{Niedermaier:2006ns,Niedermaier:2006wt,
Percacci:2007sz,Reuter:2012id,Litim:2011cp}. In a similar spirit, UV conformal extensions of the standard model with and without gravity have received attention 
\cite{Kazakov:2002jd,Gies:2003dp,Morris:2004mg,
Fischer:2006fz,
Kazakov:2007su,Zanusso:2009bs,
Gies:2009sv,Daum:2009dn,Vacca:2010mj,Folkerts:2011jz,
Bazzocchi:2011vr,Gies:2013pma,Antipin:2013exa,Dona:2013qba,Bonanno:2001xi,Meissner:2006zh,Foot:2007iy,Hewett:2007st,Litim:2007iu,Shaposhnikov:2008xi,
Shaposhnikov:2008xb,Shaposhnikov:2009pv,Weinberg:2009wa,
Hooft:2010ac,
Hindmarsh:2011hx,Hur:2011sv,Dobrich:2012nv,Tavares:2013dga,
Tamarit:2013vda,Abel:2013mya,Antipin:2013bya,Heikinheimo:2013fta,
Gabrielli:2013hma,Holthausen:2013ota,Dorsch:2014qpa,Eichhorn:2014qka,Sannino:2014lxa,Nielsen:2015una,Codello:2016muj}.
} the theories are fundamental according to the Wilsonian definition and are therefore {\it safe} from any UV cutoff. In the asymptotically free case we will investigate whether an interacting infrared fixed point  exists. When relevant we will also determine the ${\tilde a}$-function \cite{Cardy:1988cwa,Osborn:1989td,Jack:1990eb,Antipin:2013pya} at the fixed point and check the $\tilde a$-variation. 

We consider scalar extensions of the two time-honored chiral gauge theories \cite{Ball:1988xg,Appelquist:2000qg}; the generalized Georgi-Glashow (GG) \cite{Georgi:1974yf} and the Bars-Yankielowicz (BY) theories \cite{Bars:1981se} (see tables  \ref{tab:ggFields} and \ref{tab:byFields} respectively). These are both SU($N$) theories with fermions in the fundamental representation, and fermions in the two-index anti-symmetric (symmetric) representation in the GG (BY) model. Besides grand unified theories \cite{Georgi:1974yf} these theories have been employed to endow masses to standard model fermions in composite extensions of the standard model \cite{Raby:1979my} with the most recent attempt provided in \cite{Cacciapaglia:2015yra}. 

We will go beyond earlier  investigations  \cite{Appelquist:2000qg} and more recent investigations \cite{Shi:2016wnm,Shi:2015fna} by adding to the dynamics two distinct kinds of scalar matter fields; one transforming in the fundamental representation of the gauge group and one gauge singlet transforming in the bi-fundamental representation of the global symmetry\footnote{It is worth stressing that for the asymptotically safe scenario in perturbation gauge as well as Yukawa interactions are crucial for its possible existence as first argued in \cite{Litim:2014uca} and further investigated in \cite{Bond:2016dvk}.}. We will be investigating in steps first the gauge-Fermion theory that features only a gauge coupling, and then we will be considering in turn the various scalars that further induce Yukawa interactions and scalar self-interactions. We will determine the infrared trustable fixed-point dynamics for the (complete) asymptotically free theories as well as the potential emergence of interacting UV fixed points in all couplings referred to as complete asymptotic safety when asymptotic freedom is lost, extending the work of \cite{Litim:2014uca} to chiral gauge theories.

 \begin{table}[h!]
  \[ \begin{array}{c|c|cc|cc} \hline \hline
    {\rm Fields} & \left[\mathrm{SU}(N)\right] &  \mathrm{SU}(N-4+p) & \mathrm{SU}(p) & \mathrm{U}_1(1) & \mathrm{U}_2(1) \\ \hline 
      A & \asBox & 1 & 1 & N-4 & 2p\\
      \tilde F & \overline\Box & \overline\Box & 1 & -(N-2) & -p\\
      F & \Box & 1 & \Box & N-2 & -(N-p)\\
    \hline
       M & 1 & \Box & \overline\Box & 0 & N \\
      H & \overline\Box & 1 & 1 & 2 & -p \\       \hline \hline \end{array}%
  \]%
\caption{Transformation properties of the generalized Georgi-Glashow fields under the gauge and anomaly-free global symmetries.}%
\label{tab:ggFields}%
\end{table}

 \begin{table}[h!]
  \[ \begin{array}{c|c|cc|cc} \hline \hline
    {\rm Fields} & \left[\mathrm{SU}(N)\right] &  \mathrm{SU}(N+4+p) & \mathrm{SU}(p) & \mathrm{U}_1(1) & \mathrm{U}_2(1) \\ \hline 
      S & \sBox & 1 & 1 & N+4 & 2p\\
      \tilde F & \overline\Box & \overline\Box & 1 & -(N+2) & -p\\
      F & \Box & 1 & \Box & N+2 & -(N-p)\\
    \hline
      M & 1 & \Box & \overline\Box & 0 & N \\
      H & \overline\Box & 1 & 1 & -2 & -p \\
       \hline \hline \end{array}%
  \]%
\caption{Transformation properties of the generalized Bars-Yankielowicz fields under the gauge and  anomaly-free global symmetries.}%
\label{tab:byFields}%
\end{table}

The theories under investigation are built on the foundation of the chiral Lagrangian
\ea{
  \mc L_{\chi GT} &= -\frac{1}{4}F^{\mu\nu}F_{\mu\nu} + iT\sigma^\mu D_\mu\bar{T} + i\tilde{F_j}\sigma^\mu D_\mu\bar{\tilde{F}}^j + iF_k\sigma^\mu D_\mu\bar{F}^k,
}
where we have suppressed the gauge indices. The flavor indices are $j=1,2,\ldots,(N\pm4+p)$, and $k=1,2,\ldots,p$. The fermionic field $T$ refers to either $A$ or $S$ and transforms in the 2-index antisymmetric or 2-index symmetric representation of the gauge group respectively. $F$ and $\tilde{F}$ transform in the fundamental representation of the gauge group. 

We learn that it is possible to achieve complete asymptotically free chiral gauge field theories with scalars  and further, that these theories possess an infrared conformal window. 

 Once asymptotic freedom is lost in the gauge coupling, by varying the number of vector-like species, asymptotic safety can occur in gauge-Fermion theories only non-perturbatively and above a critical number of flavours.  In the presence of scalar singlets  the induced Yukawa interactions help taming the ultraviolet behaviour of the gauge interactions and perturbative asymptotic safety emerges similarly to the case of purely vector-like theories \cite{Litim:2014uca}. 
 
 Our results extend the number of theories that can be fundamental according to Wilson \cite{Wilson:1971bg,Wilson:1971dh} to the case of chiral gauge theories with scalars. In fact the occurrence of UV complete fixed points guarantees the fundamentality of the theory since, setting aside gravity, it means that the theory is valid at arbitrary short distances \cite{Wilson:1971bg,Wilson:1971dh}.   

\section{Gauge-fermion analysis of the BY and GG generalised theories}\label{sec:gaurge-fermion}
We begin by re-examining and extending the investigations of the conformal dynamics of BY and GG theory without scalars. To enable us to easily compare our analysis across different values of the number of colors, $N$, we will replace $p$ by $x=p/N$ in the much of following, and keep in mind that the theory is only physical for certain values of $x$. The beta function to three loop order can be found in the appendix \eqref{eq:by-gg-fermion-beta}. We note that in the limit of large $N$ and large $p$ with the ratio $x=\frac{p}{N}$ held constant (which we will refer to as the Veneziano limit), the BY and GG theories have the same beta functions, and indeed it can be shown that the theories are completely equivalent in this limit.

In our search for fixed points, will use the Banks-Zaks method, where we start out by finding the value of $x$ where the one loop term in the beta function vanishes for a given $N$ and call this $x_{AF}$. For $x>x_{AF}$ the theory is infrared free and for $x<x_{AF}$ the theory is asymptotically free. We have  \ea{
  x_{AF} &= \frac{9}{2}\mp\frac{3}{N}.
}

\subsection{Asymptotically Free Dynamics and Conformal Window}
We first investigate the phase diagram for the  asymptotically free regime of the theory.
\subsubsection{Veneziano limit}
In this limit the ratio $x = p/N$ is held constant and we rescale the coupling by $N$ as follows
\ea{
 \bar a_g    = \frac{g^2 N}{(4\pi)^2}.
}
\begin{figure}[bt] 
\includegraphics[width=.45\columnwidth]{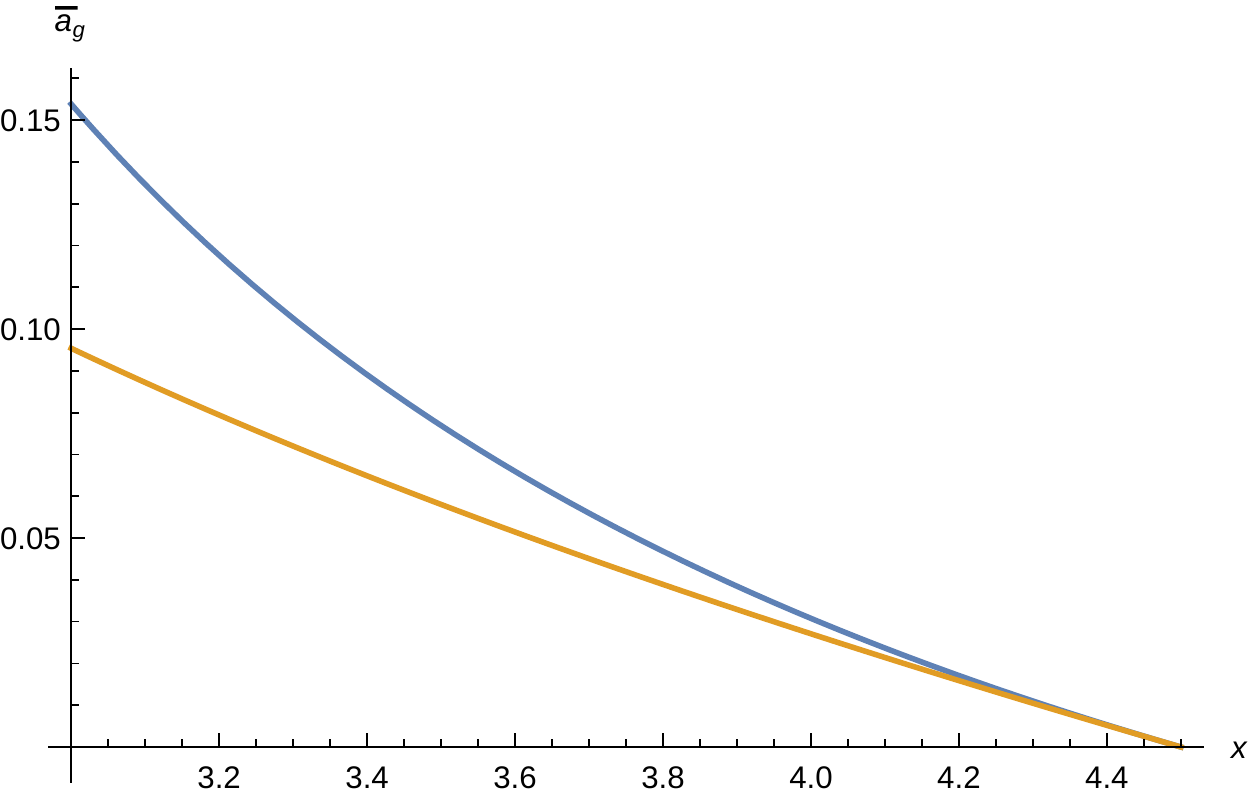}
  \caption{Fixed point values of the gauge-fermion theory in the Veneziano limit. The blue line is the well-known 2 loop result, and the yellow our improved 3 loop one.}\label{fig:vLimitFG1}\end{figure}
For convenience, we write the beta function in this case explicitly.
\ea{
  \beta_{\bar a_g} &= -\bar a_g^2 \left(6-\frac{4 x}{3}\right) - \bar a_g^3 \left(13-\frac{26 x}{3}\right) - \bar a_g^4 \left(\frac{127}{3}-\frac{979 x}{18}+\frac{112 x^2}{27}\right),
}

 Here the Banks-Zaks fixed point is an IR one. It is found by setting $\beta_{\bar a_g}=0$ and by picking the solution which vanishes smoothly for $x=x_{AF}$, as is seen in Fig.~\ref{fig:vLimitFG1}.
 
However, the three loop term introduces a second fixed point that will be discussed later.

\subsubsection{Finite $N$ and $p$ Conformal Window}
From a phenomenological point of view it is interesting to cover also the low $N$ limit. Since the GG theory is defined only for $N\geq5$, we will use this as a reference value, but also consider the conformal window for any $N$ and $p$.

\begin{figure}[hbt]
 \subfloat[Fixed points values of the gauge-fermion BY theory with $N=5$. The blue line is the well-known 2 loop result, and the yellow our improved 3 loop one.]{\includegraphics[width=.45\columnwidth]{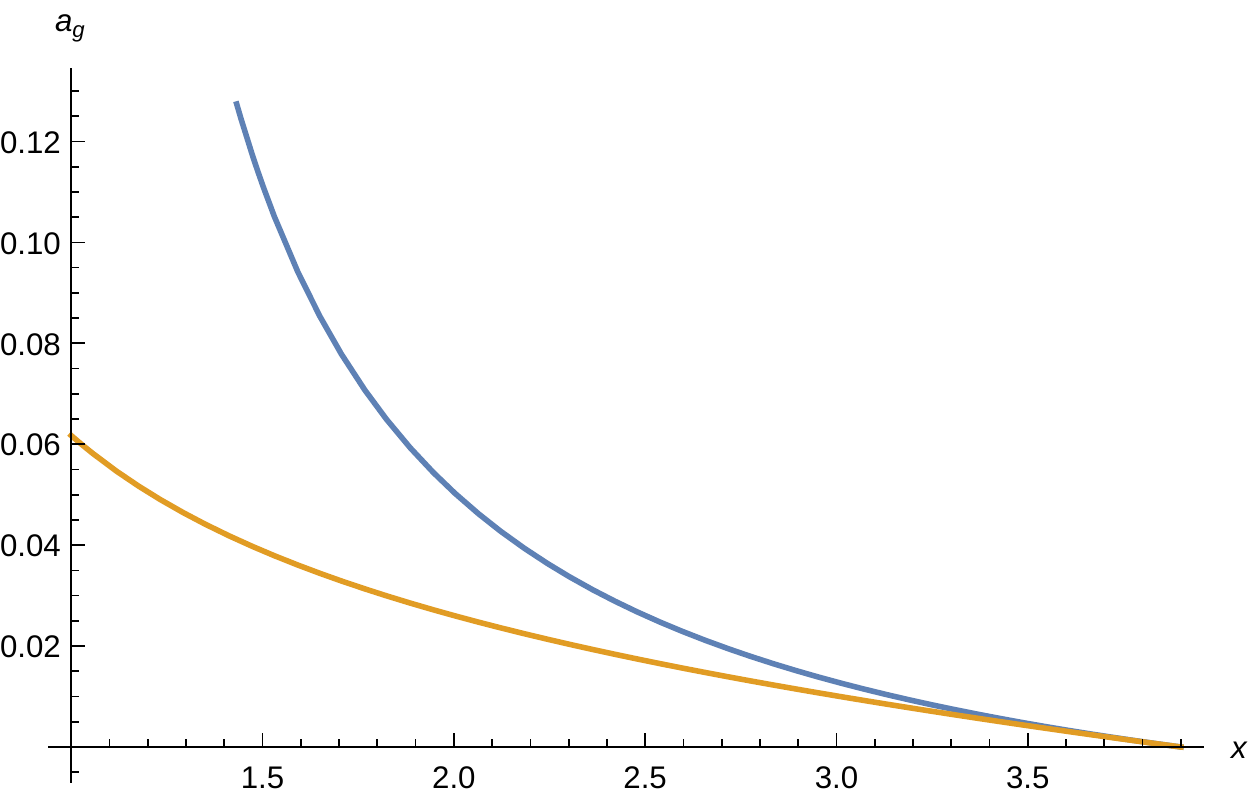}\label{fig:n5byFG1}}~~~~
   \subfloat[Fixed points values of the gauge-fermion GG theory with $N=5$. The blue line is the well-known 2 loop result, and the yellow our improved 3 loop one.]{\includegraphics[width=.45\columnwidth]{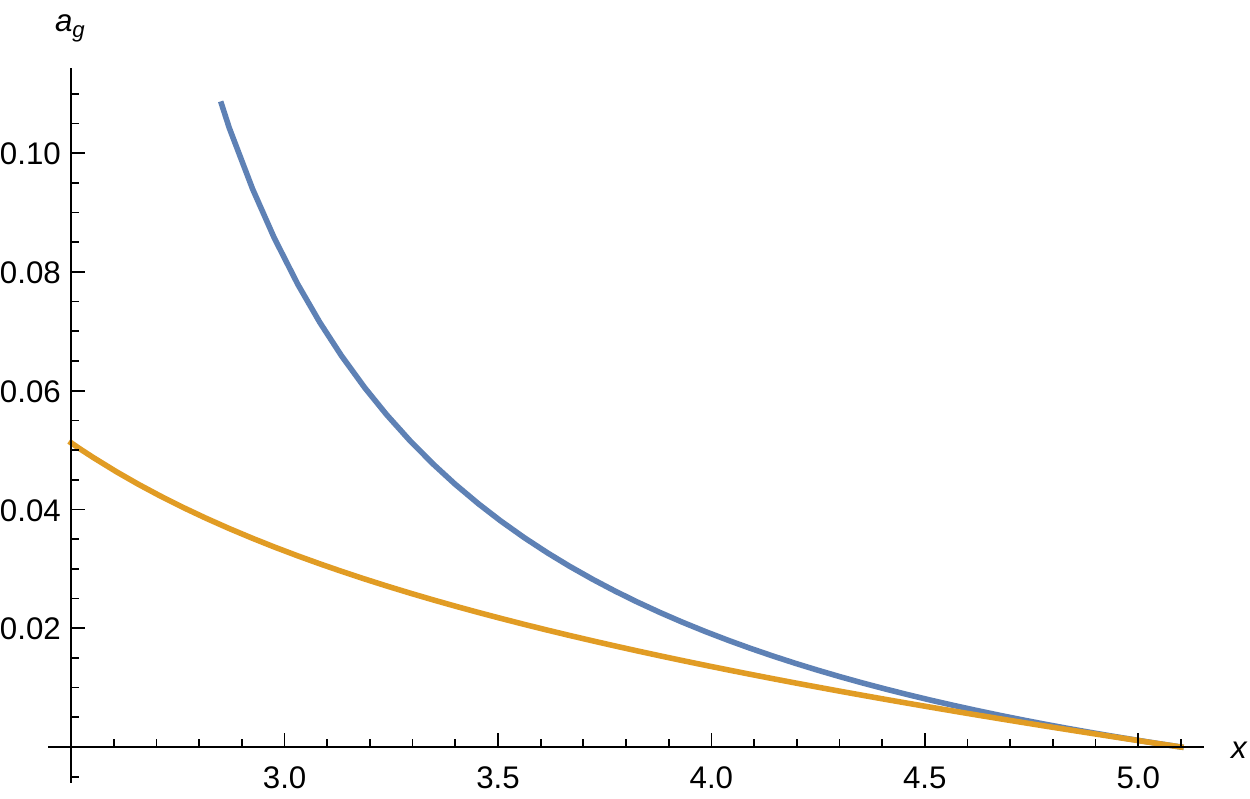}\label{fig:n5ggFG1}}
 \caption{$N=5$ in gauge-Fermion BY and GG theory.}
\end{figure}

For $N=5$ we proceed exactly as in the Veneziano case above.  Here we have that BY and GG possess a qualitatively similar picture, see Figs. \ref{fig:n5byFG1} and \ref{fig:n5ggFG1}. A similar picture is also found for the BY model for $N=2$, but since the GG theory cannot be extended to such low values, we do not discuss it further.

\begin{figure}[hbt]
  \includegraphics[width=.85\columnwidth]{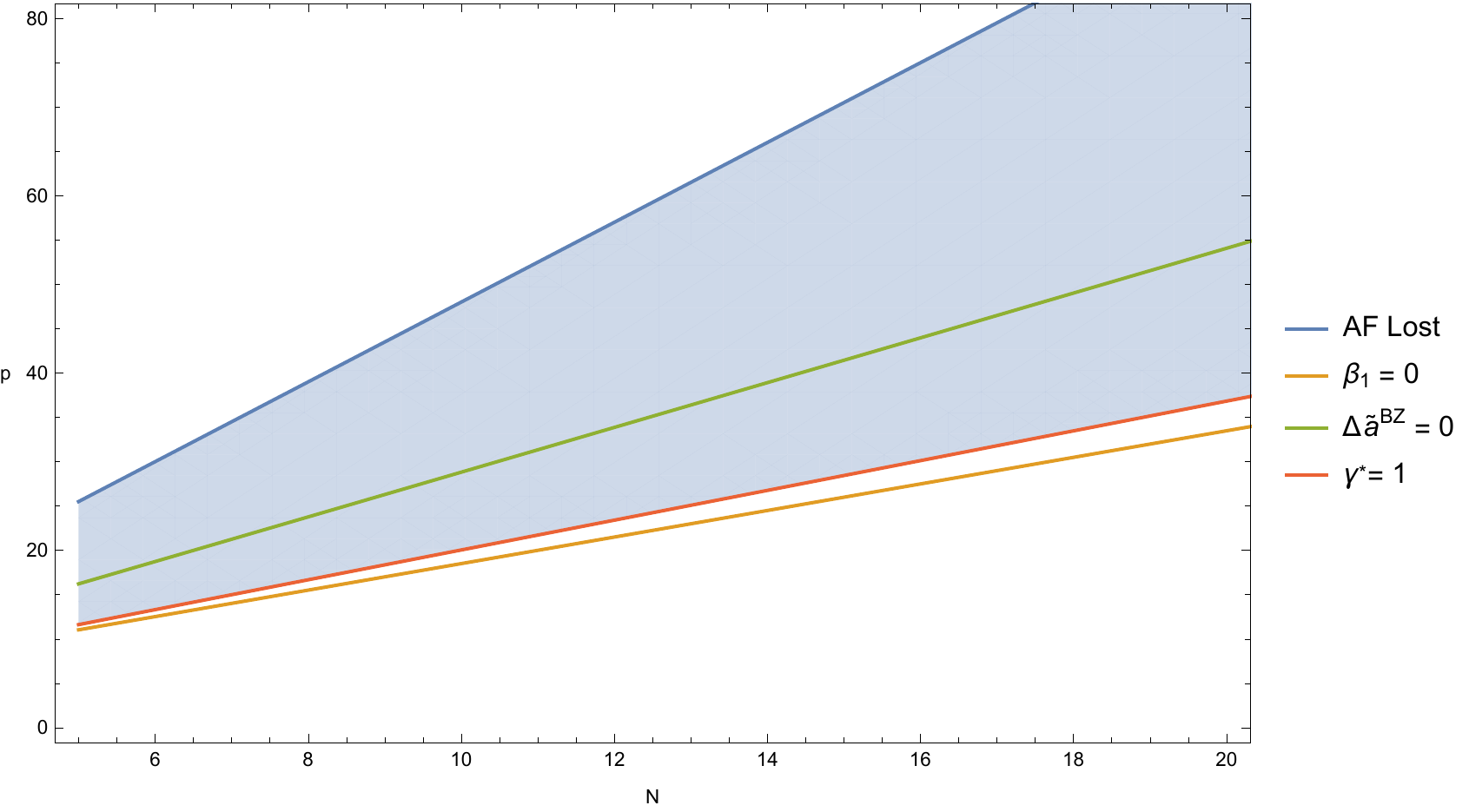}
  \caption{The conformal window for $N \geq 5$ Generalized Georgi-Glashow theory. From above, the lines are: The border between IR-freedom and asymptotic freedom, $\Delta\tilde{a}^{\mathrm{BZ}}=0$, $\gamma^*=1$, and $\beta_1=0$.}\label{fig:PhaseDiagramGG}
\end{figure}

It is conventional to speak of the \emph{conformal window}, that is the region in parameter space where the theory is asymptotically free and has a trustable IR fixed point. To determine the conformal window in the theories discussed in this paper, we restore the parameter $p$ and work in the parameter space spanned by $N$ and $p$. The upper boundary of the conformal window is uniquely given by the line for which the one loop beta function vanishes
\ea{
  \beta_0&=-2-3N+\frac{2}{3}p=0\\
  p_{AF}&= \frac{3}{2}(2+3N).
}
For definitiveness we consider explicitly the conformal window for the GG theory since the one for the BY theory is similar. To estimate the lower boundary of the conformal window we use several methods. One could simply ask when the two loop beta function ceases to have a fixed point, which happens when the two loop term vanishes, $\beta_1=0$. However, at this point, the putative fixed point value diverges, indicating that perturbative control has long been lost. Another method, which draws upon our non-perturbative knowledge of the theory, is to define the limit as the point where the anomalous dimension of the fermion mass operator at the fixed point equals two, $\gamma^*=2$. For anomalous dimensions larger than two, the associated scalar operator would violate the unitarity bound \cite{Mack:1975je}. Instead of this method, we will use the more conservative expectation that the lower boundary of the conformal window occurs for $\gamma^*$ around unity when four-fermion operators cannot be neglected since they can drive chiral symmetry breaking. Yet a fourth possibility \cite{Antipin:2013qya} is to insist that, along the flow connecting the IR and UV fixed points, the $\tilde{a}$-function of Osborn \cite{Osborn:1989td,Jack:1990eb} has the property  \cite{Cardy:1988cwa} that 
\ea{
  \Delta\tilde{a} = \tilde{a}^{UV}-\tilde{a}^{IR} \geq0.
}
This inequality was conjectured by Cardy \cite{Cardy:1988cwa}, it has been show to hold in the limit of vanishing coupling constants \cite{Osborn:1989td,Jack:1990eb}, and it has since been argued to hold non-perturbatively \cite{Komargodski:2011vj,Komargodski:2011xv}. 
We consider here all these estimated lower boundaries and note that they each give different constraints  with the most constraining coming from the perturbative positivity of $\Delta \tilde{a}$.  We present the conformal window for the generalized Georgi-Glashow theory in figure~\ref{fig:PhaseDiagramGG}. 

Alternative nonperturbative suggestions to estimate the lower boundary of the conformal window and the possible infrared phases of these theories have been discussed in \cite{Appelquist:2000qg}.

Finally the conformal window for the BY theory at large $N$ agrees with the GG one by construction while qualitatively is very similar to the GG at smaller $N$. 

\subsection{Asymptotically Safe Conformal Window without Scalars}
\begin{figure}[tb]
   \subfloat[Fixed points values of the gauge-fermion theory in the Veneziano limit. The blue line is the three loop Banks-Zaks-like fixed point that moves to negative values once asymptotic freedom is lost. The yellow shows the emergence of a non-Banks-Zaks asymptotically safe FP.]{\includegraphics[width=.45\columnwidth]{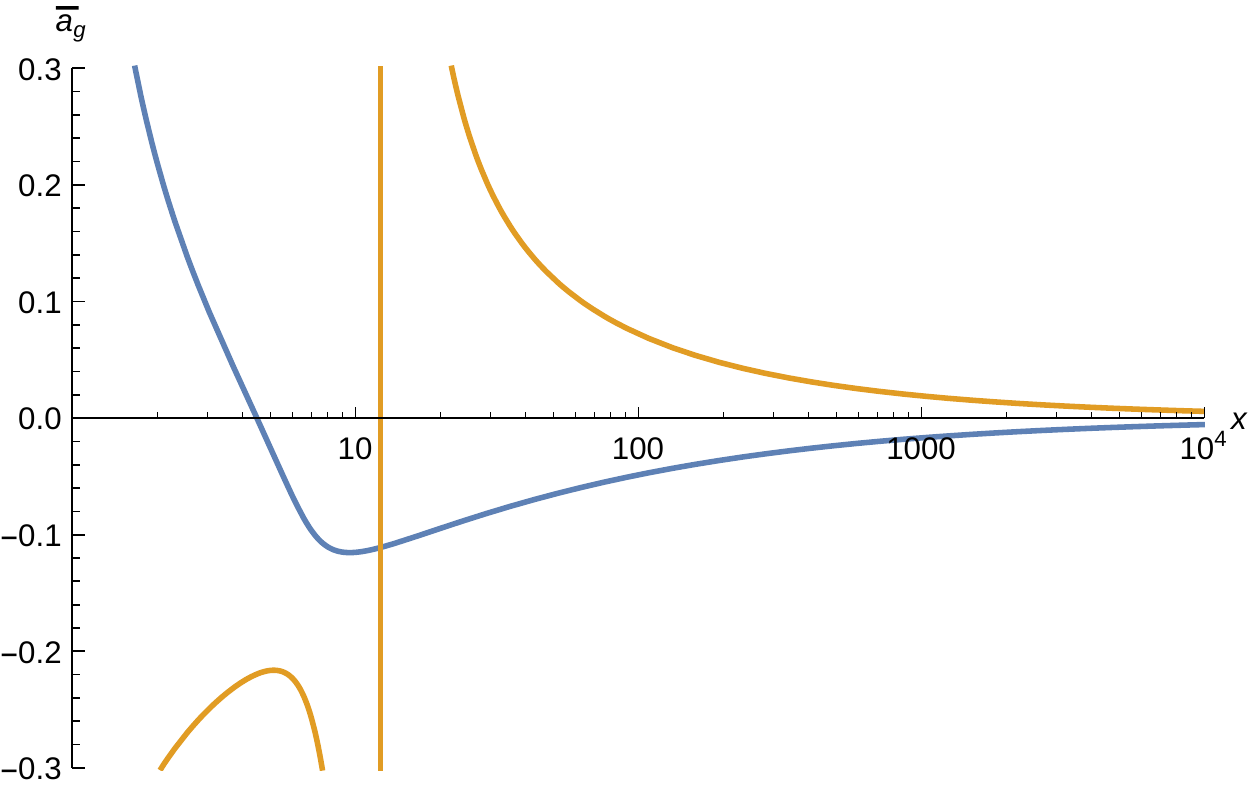}\label{fig:vLimitFG2}} ~~~~~~~
  \subfloat[Fixed points values of the gauge-fermion GG theory with $N=5$. The blue line is the  three loop Banks-Zaks-like fixed point that moves to negative values once asymptotic freedom is lost. The yellow shows the emergence of a non-Banks-Zaks asymptotically safe FP.]{\includegraphics[width=.45\columnwidth]{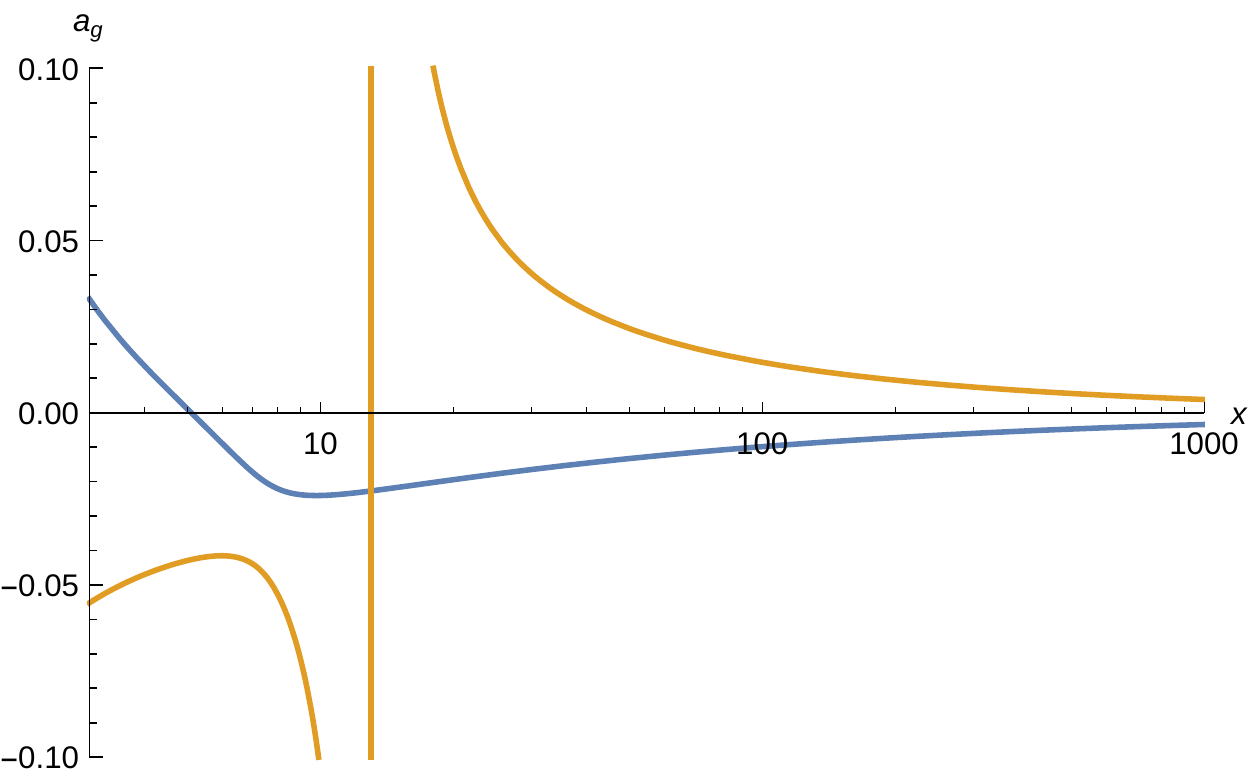}\label{fig:n5ggFG2}}
  \caption{Three-loops Asymptotically Safe Fixed point in the Veneziano limit and for the $N=5$ GG gauge-Fermion theory.  }
\end{figure}

For $x>x_{AF}$ the theory is  infrared free and develops a Landau pole at one loop. At two loops and in the trustable perturbative regime it does not develop an interacting UV fixed point in agreement with the results of \cite{Caswell:1974gg}. This theory, however, might still become asymptotically safe in the large $p$ limit in a fashion similar to the one investigate in \cite{Pica:2010xq} for a purely vector-like theory. In fact, a tantalising hint that asymptotic safety can indeed emerge here is provided by a careful analysis of the three-loops results. Here we observe the occurrence of an interacting UV fixed point with the  coupling value at criticality that decreases as we increase the number of vector-like fermions $p$.  The value of the UV fixed point coupling both in the Veneziano limit and GG theory for $N=5$  (the BY theory has an equivalent behaviour) is shown respectively in Figure \ref{fig:vLimitFG2} and \ref{fig:n5ggFG2} as function of $x$. The blue curve is the three loop result for the Banks-Zaks fixed point that once asymptotic freedom is lost moves to the negative axis and becomes unphysical. The yellow curve shows the emergence of an asymptotically safe non-Banks-Zaks-like fixed point when asymptotic freedom is lost and $x$, i.e. the number of flavours, is above a critical value. 

\begin{figure}[hbt]
  \includegraphics[width=.85\columnwidth]{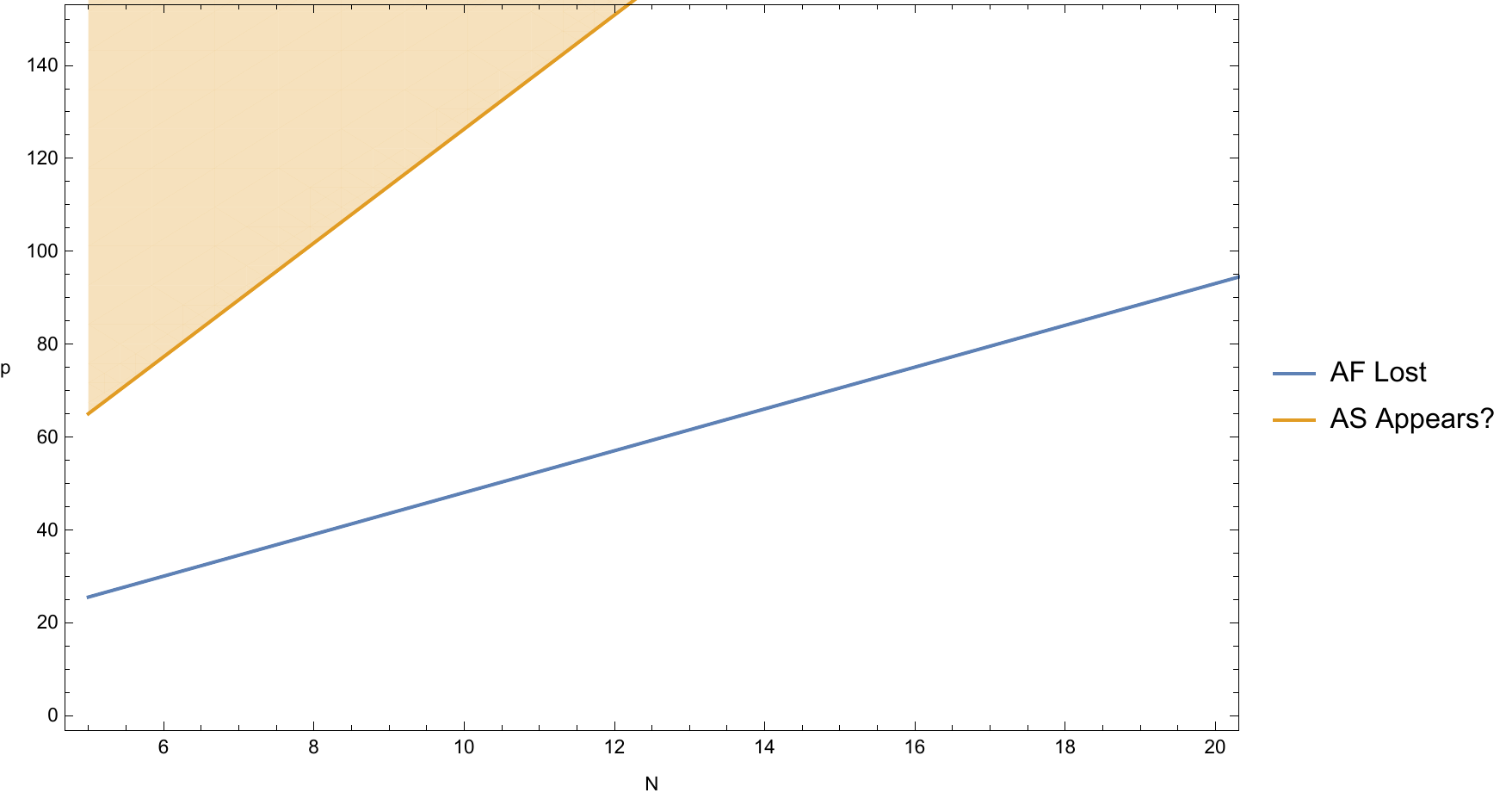}
  \caption{Asymptotically safe conformal window for the Generalized Georgi-Glashow theory via the three loop order estimate.  The shaded region is the area where the asymptotic safety sets in.   The blue line marks the loss of asymptotic freedom. }\label{fig:PhaseDiagramGG-AS}
\end{figure}

The potentially novel asymptotically safe conformal window is shown in Figure \ref{fig:PhaseDiagramGG-AS}. The qualitative feature of this asymptotically safe window is that it would start at a critical number of flavors above the loss of asymptotic freedom and would then continue for any number of flavors above that. Of course, because of the absence of a perturbatively trustable Banks-Zaks-like fixed point this picture needs independent confirmation. It is in line, however, with similar expectations at large number of flavors in vector-like theories discussed in \cite{Holdom:2010qs,Pica:2010xq}. 

If asymptotic safety were to occur in these theories, like for the vector-like case  \cite{Holdom:2010qs,Pica:2010xq}, because of the absence of a Bankz-Zaks fixed point a critical number of flavours must necessarily develop such that in between the loss of asymptotic freedom and this value the theory cannot be fundamental. Above this critical value the theory admits a continuum limit. The crucial fact is that these theories could become asymptotically safe  because of the sufficiently large number of fermions rather than due to the balancing effect of Yukawa interactions in theories featuring also scalars \cite{Litim:2014uca}. In these theories scalars would not be needed to restore the fundamentality of the theory when asymptotic freedom is lost.

To elucidate the question of whether this putative fixed point is indeed physical or a mere artifact of perturbation theory, we have computed $\Delta\tilde a^{NBZ}$, the change in the $\tilde a$-function between the ultraviolet non-Banks-Zaks fixed point and the infrared Gaussian fixed point, and for all relevant values of $p$ and $N$, we find that it is negative which appears to be a strike against the perturbative trustability of this fixed point. We also find that the anomalous dimension of the $\tilde F F$ operator at this fixed point is always negative. Of course, these results imply that non-perturbative methods must be considered here to decide whether a new UV-safe conformal window emerges. 
%
%
%
%

%

\FloatBarrier
\section{Generalized chiral gauge theories with a meson-like scalar}

We now move to consider chiral gauge theories that include also scalars and investigate their phase structure. Because of the presence of scalars, new interactions become possible such as Yukawa and self-interactions.  This means that new marginal couplings need to be considered including their beta functions. We provide the detailed analysis for the examples that we found most representatives and comment on the general results later. 

 We start by adding a mesonic-like scalar field 
  $M$ which is a singlet under the SU($N$) gauge group, and bifundamental under the global $\mathrm{SU}(N\pm4+p)\times\mathrm{SU}(p)$ group. This means that the Lagrangian will be extended to include Yukawa interactions and scalar self-interaction and assume the generic form:
\ea{
  \mc{L} &= \mc L_{\chi GT} + \mc L_M  \label{Lagrangian-meson}\\
  \mc L_M &= \Tr[\p_\mu M^\dagger\p^\mu M]+(y_M \tilde F_j M^j_k F^k + h.c.) + u_1 \Tr[M^\dagger M]\Tr[M^\dagger M] + u_2 \Tr[M^\dagger M M^\dagger M].
}
The newly introduced coupling constants are rescaled as follows
\ea{
  a_M = \frac{y_M^2}{(4\pi)^2}  \ ,\qquad z_1 = \frac{u_1}{(4\pi)^2} \ , \qquad z_2 = \frac{u_2}{(4\pi)^2}  \ , 
}
and the full set of beta functions are given in equations \eqref{eq:xGT-beta-ag}-\eqref{eq:xGT-beta-z2}. 
 Because the newly introduced scalar does not modify the one-loop gauge beta function asymptotic freedom for the gauge coupling is lost again for 

\ea{
  x_{AF} &= \frac{9}{2}\mp\frac{3}{N}.
}

We now investigate the IR conformal dynamics of this theory both in the Veneziano and finite $N$ and $p$ limits. 

\subsection{Complete Asymptotic Freedom in the Veneziano limit}
In this limit the two theories are degenerate and the double trace coupling $z_1$ decouples from the running of the other couplings. The opportunely rescaled couplings read
\ea{
   \bar a_M = \frac{y_M^2N}{(4\pi)^2} \ ,\qquad   \bar z_1 = \frac{u_1 p^2}{(4\pi)^2}  \qquad \, \bar z_2 = \frac{u_2 p}{(4\pi)^2} \ .
}

Because of the presence of Yukawa and scalar self-coupling interactions, this theory does not in general allow for a continuum limit, even when the gauge coupling is asymptotically free. One has to further study the one-loop conditions for the Yukawa and scalar self-coupling interactions to ensure that they are also asymptotically free. Since, at least at large $N$, the double-trace operator is a spectator coupling, the general conditions for this to happen reduces to the ones presented  in \cite{Pica:2016krb} that we review here for the reader's convenience. In the parameter space region where  the Yukawa coupling $\bar a_M$ vanishes faster than $\bar a_g$, the conditions for complete asymptotic freedom are
\ea{
  b_0 < 0, \quad b_0-c_1>0,\quad k\geq0,\quad b_0-d_2+\sqrt{k}>0,
  \label{CAF1}
}
with these coefficients related to the beta functions via 
\begin{eqnarray}
\beta_g &=&  \bar a_g^2 \left(b_0 + b_1 \bar a_g + b_M \bar a_M \right) \\
\beta_M &=& \bar a_M \left( c_1 \bar a_g + c_2 \bar a_M \right) \\
\beta_{z_2} &=& \bar z_{2} \left( d_1\bar z_{2} + d_2 \bar a_g + d_3 \bar a_M \right) + d_4 \bar a_g^2 + d_5 \bar a_M^2
\end{eqnarray}
and 
\begin{eqnarray}\label{eq:k}
k &=&  (b_0 - d_2)^2 - 4 d_1d_4 
\end{eqnarray}
For the theory studied here, and within the regime of interest, we have: 
\ea{
  b_0 = \frac{4x}{3}-6, \quad b_0-c_1=\frac{4x}{3},\quad k=\left( \frac{4x}{3}-6\right)^2 , \quad b_0-d_2+\sqrt{k}=0,
}
where in the last equation we have used that $b_0<0$. Thus, the first three conditions are satisfied when $0<x<\frac{9}{2}$, and the last in \eqref{CAF1} fails to be satisfied for all values of $x$ since $d_2=d_4=0$.

Along the fixed flow given by $\bar a_y = \frac{c_2}{b_0-c_1}\bar a_g$ (see \cite{Pica:2016krb} for details), the conditions are 
\ea{
  b_0 < 0, \quad b_0-c_1>0,\quad k'\geq0,\quad b_0-d_2'+\sqrt{k'}>0,
}
with 
\begin{equation}
  d_2' = d_2 +d_3\frac{b_0-c_1}{c_2}, \qquad k' =  \left( b_0 - d_2 -d_3 \frac{b_0-c_1}{c_2}  \right)^2 - 4d_1 \left(d_4 + d_5 \left( \frac{b_0-c_1}{c_2} \right)^2 \right) \label{eq:k'}
\end{equation}
and we find
\ea{
  k'=\left(6-\frac{4x}{3}+\frac{16x}{3(3+2x)}\right)^2+\frac{512x^2(1+2x)}{9(3+2x)^2},\quad b_0-d_2'+\sqrt{k'}=-6+\frac{4x}{3}-\frac{16x}{3(3+2x)}+\sqrt{k'}>0.
}
which satisfies the conditions for $0<x<\frac{9}{2}$. 

Since the final condition of \eqref{CAF1} only fails to be satisfied when the influence of the Yukawa coupling is ignored entirely, we interpret these results as complete asymptotic freedom being found for all values of $a_g$ and $a_y$ in the region bounded by the fixed-flow line and $a_y=0$. 
 
\begin{figure}[hbt]
 \subfloat[RG flow of the Veneziano limit of the generalized GG/BY theory with mesons. This slice of parameter space has $x=4.2$ and $\bar z_2=\bar z_2^\star\approx0.1293$.]{\includegraphics[width=.30\columnwidth]{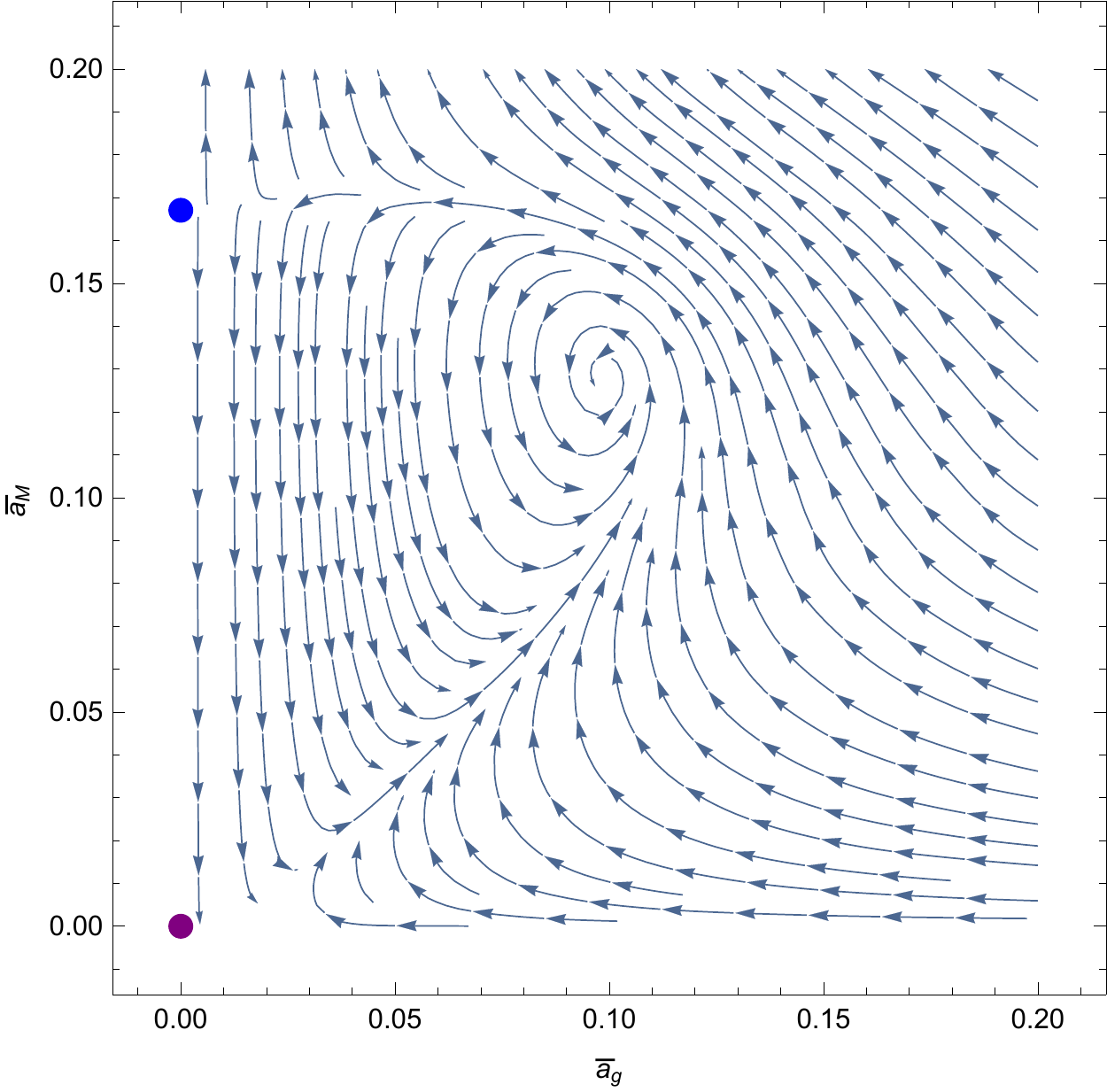}\label{fig:caf-flow-ggM-FP1}}\hfill
 \subfloat[RG flow of the Veneziano limit of the generalized GG/BY theory with mesons. This slice of parameter space has $x=4.2$ and $\bar z_2=\bar z_2^*\approx0.0299$.]{\includegraphics[width=.30\columnwidth]{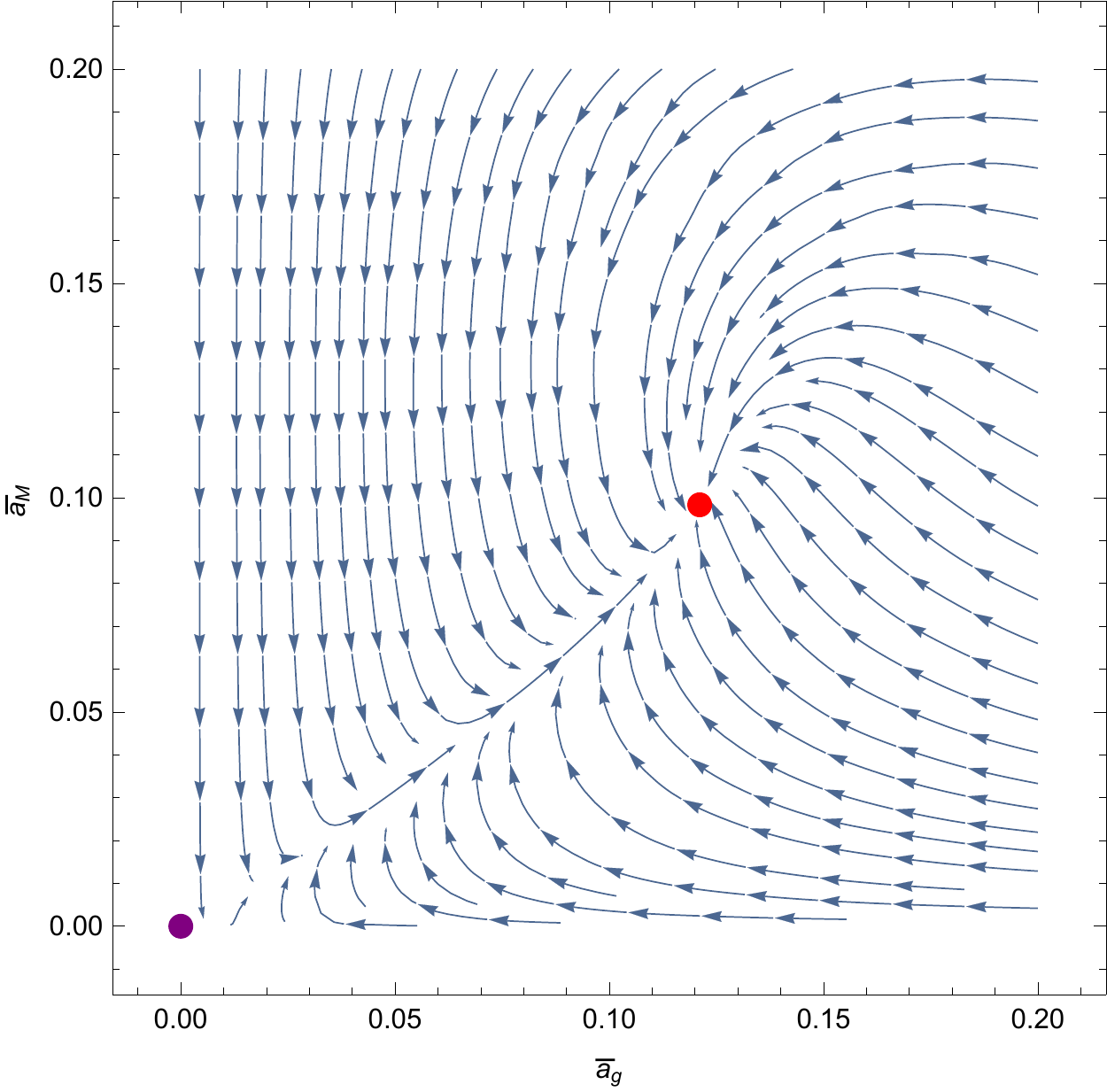}\label{fig:caf-flow-ggM-FP2}}\hfill
 \subfloat[RG flow of the Veneziano limit of the generalized GG/BY theory with mesons. This slice of parameter space has $x=4.2$ and $\bar z_2=0$.]{\includegraphics[width=.30\columnwidth]{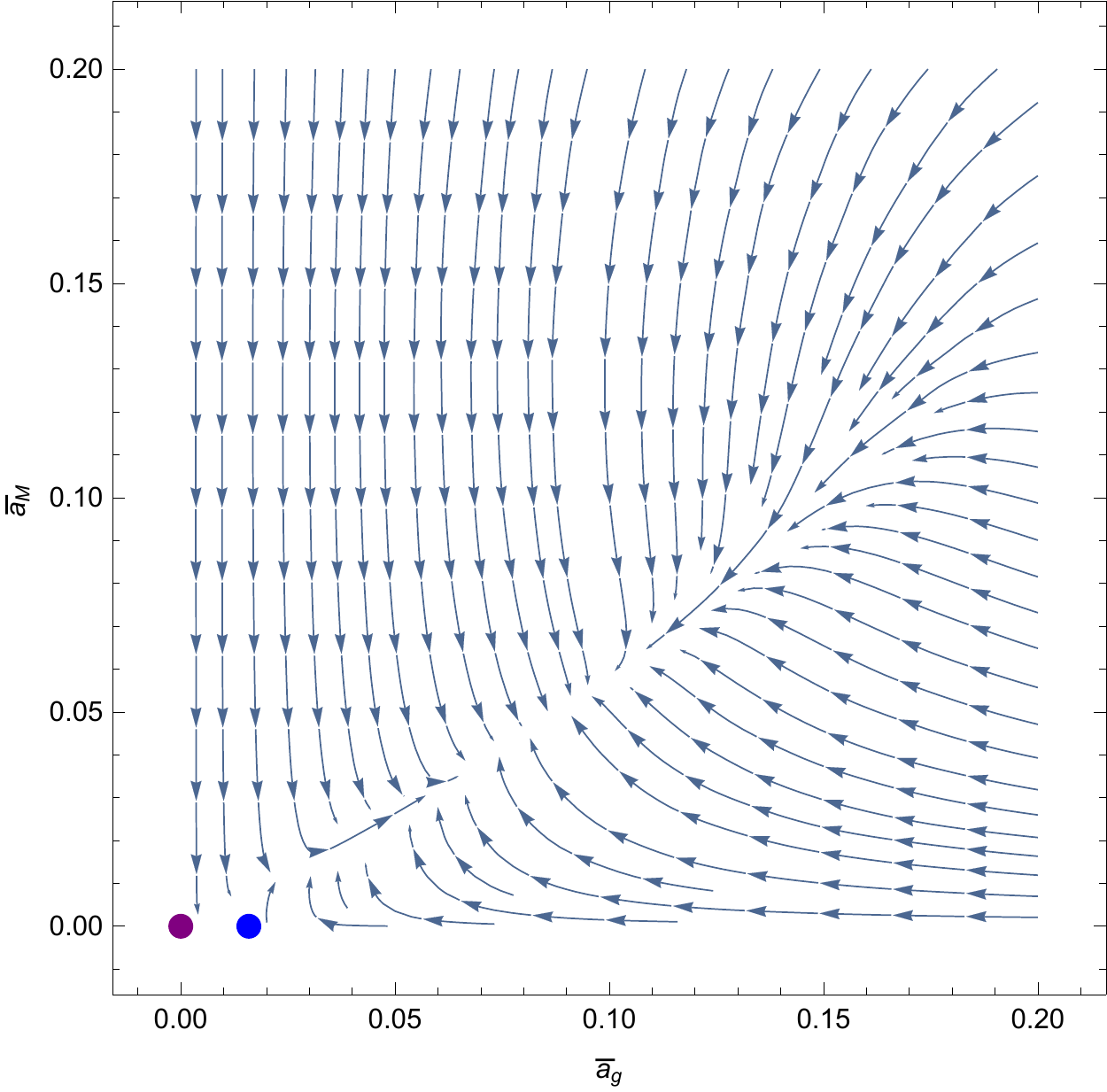}\label{fig:caf-flow-ggM-FP3}}
 \caption{Three slices of the $\bar a_g-\bar a_M$ paramter space. We see that there is an all-directions IR-attractive fixed point in figure b). It along with the two fixed points in figure c) indicate the boundaries of the region of complete asymptotic freedom where all flows go to the Gaussian fixed point in the UV. The points marked in purple, red, and blue are the fixed points of the theory with purple being the Gaussian fixed point, red the all-directions IR-attractive fixed point, and blue the additional fixed points.}\label{fig:CAFGG}
\end{figure}

We present in Fig.~\ref{fig:CAFGG} the renormalization group (RG) flow for pairs of couplings demonstrating the existence of a completely asymptotically free region, as well as the IR-attractive fixed points discussed in the IR dynamics paragraph.

\subsubsection{Conformal IR dynamics}
The presence of IR fixed points can be investigated  independently of the complete asymptotically free  analysis since the  RG trajectories will inevitably end at the IR fixed point. 

\begin{figure}[hbt]
  \subfloat[Values of the primary fixed point of the mesonic gauge-Yukawa theory in the Veneziano limit. There is also another value of $\bar{z}_1$ which gives a fixed point, but this is the IR stable solution.]{\includegraphics[width=.45\columnwidth]{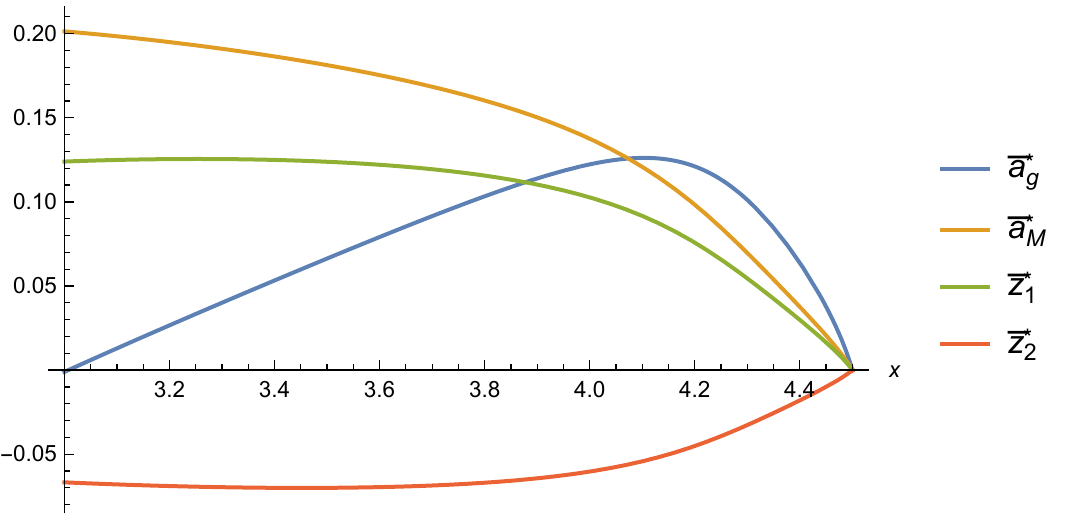}\label{fig:vLimitGY1}}
  \subfloat[Values of the secondary fixed point of the mesonic gauge-Yukawa theory in the Veneziano limit. Notice that for these values, there is no real-valued fixed point for $\bar{z}_1$.]{\includegraphics[width=.45\columnwidth]{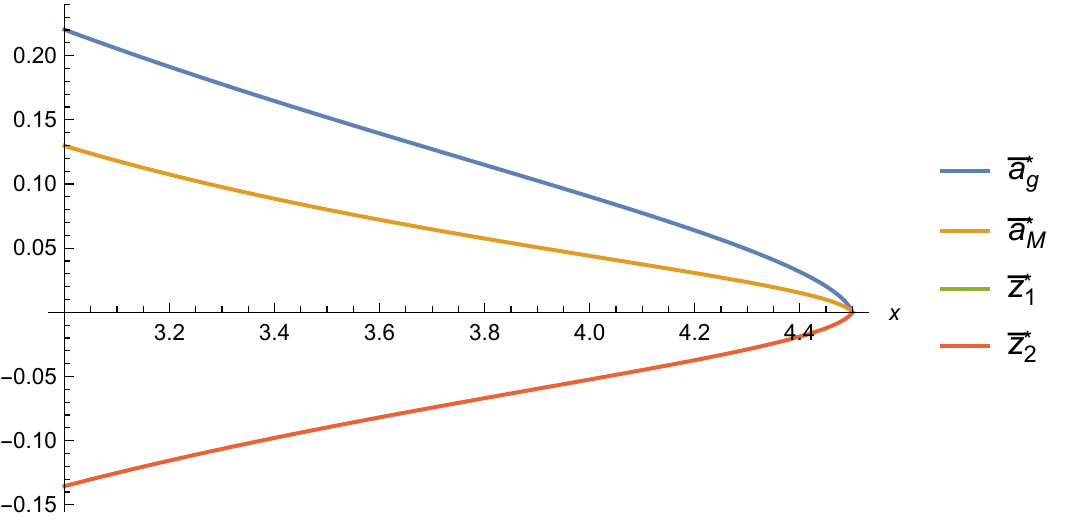}\label{fig:vLimitGY2}}
  \caption{Veneziano limit of the mesonic gauge-Yukawa theory}
\end{figure}

For $x<x_{AF}$ we have two Banks-Zaks type fixed points (meaning that they vanish at $x=x_{AF}$), one for positive $z_2$, which has two corresponding solutions for $z_1$, and one fixed point with negative $z_2$ and only imaginary solutions for $z_1$. For further details, see Figs. \ref{fig:vLimitGY1} and \ref{fig:vLimitGY2}. Since the second fixed point has a negative value for the self-coupling, the theory described by this fixed point is unstable and we will not consider it further. We refer to the fixed values of the first fixed point as $\bar a_g^*,\bar a_M^*,\bar z_1^*,$ and $\bar z_2^*$. The analysis of this fixed point  follows closely the one of the theory described in \cite{Antipin:2011aa,Antipin:2013pya} and we will only deal with it briefly here. Note that there are other fixed points which can be found by allowing $\bar a_g, \bar a_M$ or $\bar z_2$ to equal zero. We have in figure \ref{fig:caf-flow-ggM-FP1} referred to the fixed point one finds by setting $\bar a_g=0$ by $\bar a_M^\star$ and $\bar z_2^\star$.

It is interesting to note (see Figure \ref{fig:vLimitGY1}) that where the fixed point value of $a_g$ in the gauge-Fermion case diverges for low $x$, we here find that the presence of Yukawa and quartic couplings forces the fixed point value down to instead vanish at low $x$.

\subsubsection{Finite $N$}
\begin{figure}[hbt]
  \subfloat[The fixed point values when the almost-decoupled $z_1$ coupling numerically smallest.]{\includegraphics[width=.45\columnwidth]{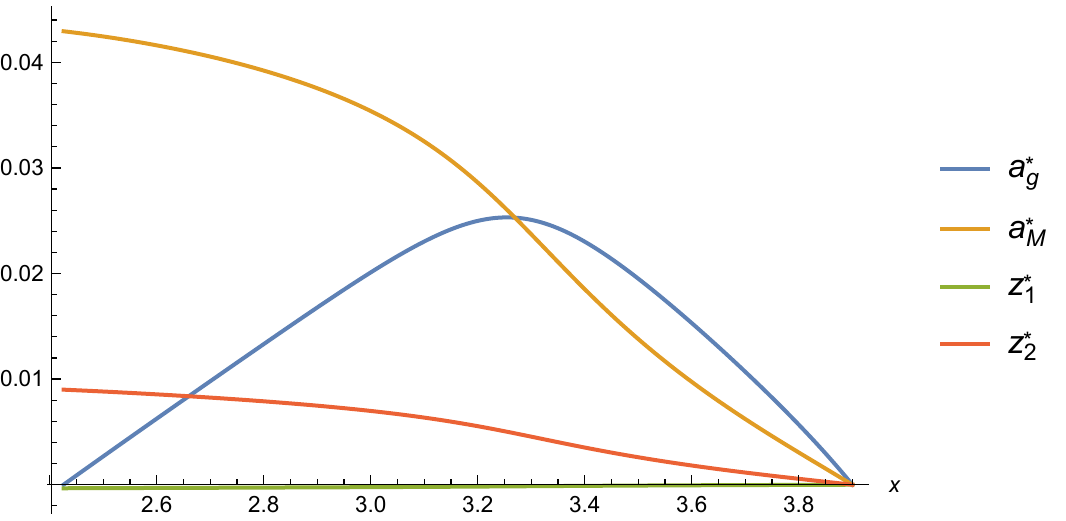}\label{fig:n5byGY1}}
  \subfloat[The fixed point values when the almost-decoupled $z_1$ coupling numerically largest.]{\includegraphics[width=.45\columnwidth]{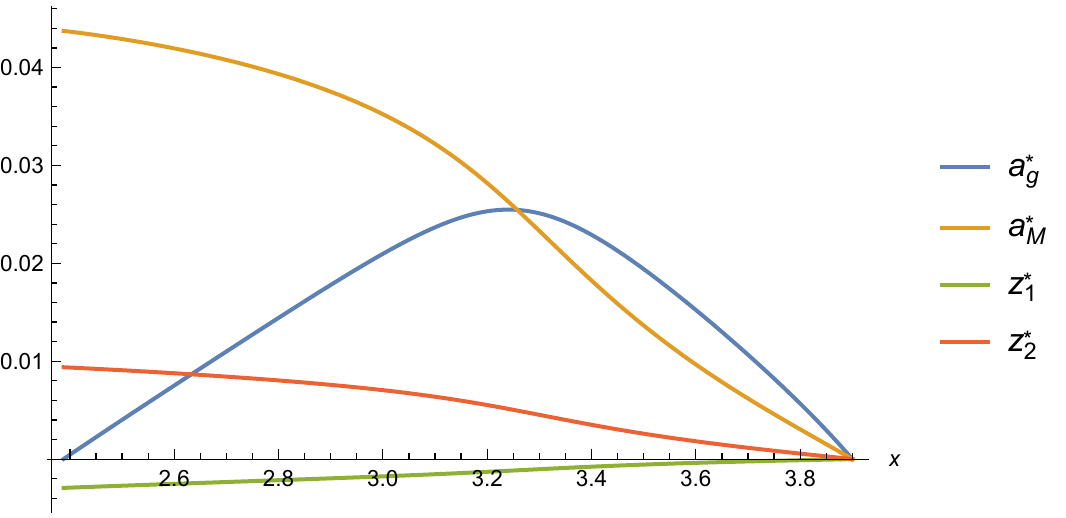}\label{fig:n5byGY2}}
  \caption{Fixed points values of the mesonic gauge-Yukawa BY theory with $N=5$.}
\end{figure}

We proceed by examining the IR dynamics of the mesonic gauge-Yukawa BY theory for $N=5$, and find that the fixed point with negative $z_2$ (corresponding to Figure \ref{fig:vLimitGY1}) has disappeared, while the one with positive $z_2$ (corresponding to Figure \ref{fig:vLimitGY2}) remains. We also see that even at finite $N$, the contribution from the double trace operator $z_1$ is small, in that the fixed point locations for $a_g,a_M$ and $z_2$ are largely unchanged. In Figs. \ref{fig:n5byGY1} and \ref{fig:n5byGY2}, we have plotted these fixed point locations, $a_g^*,a_M^*,z_2^*$ and $z_1^*$, for the two closely related fixed points.

\begin{figure}[hbt]
  \subfloat[The fixed point values when the almost-decoupled $z_1$ coupling numerically smallest.]{\includegraphics[width=.45\columnwidth]{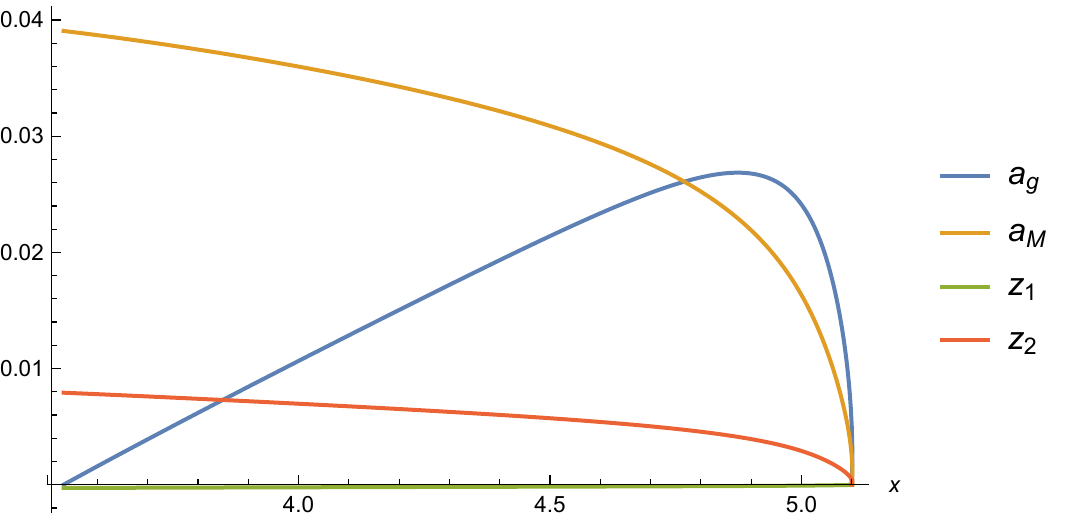}\label{fig:n5ggGY1}}
  \subfloat[The fixed point values when the almost-decoupled $z_1$ coupling numerically largest.]{\includegraphics[width=.45\columnwidth]{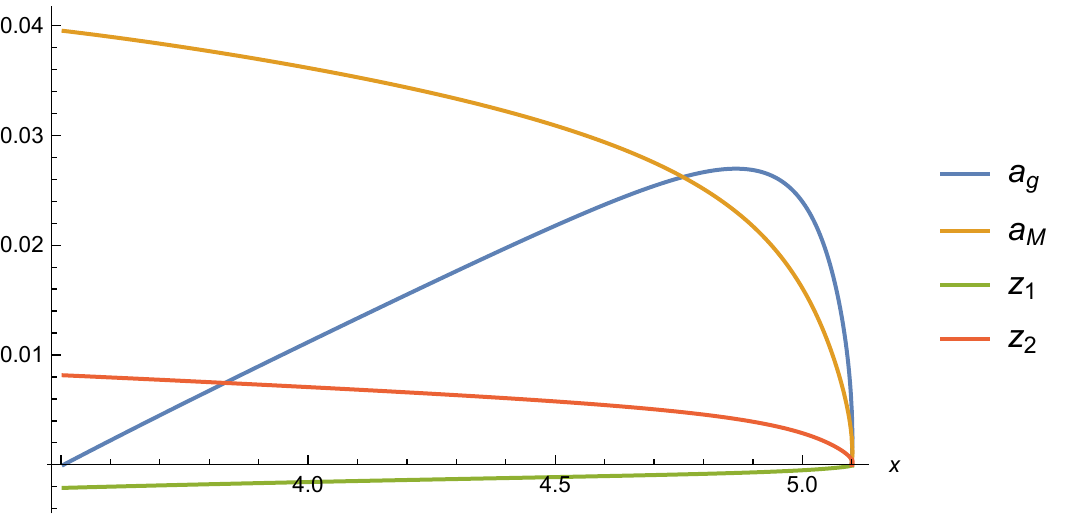}\label{fig:n5ggGY2}}
  \caption{Fixed points values of the mesonic gauge-Yukawa GG theory with $N=5$. }
\end{figure}

Moving on to the mesonic gauge-Yukawa GG theory for $N=5$, we find a very similar pattern repeated once more, see Figs  \ref{fig:n5ggGY1} and \ref{fig:n5ggGY2}. {However, careful observation will show that the fixed point values no longer vanish as the border of asyptotic freedom, $x_{AF}=\frac{51}{10}$, is approached from below. We will return to this point in the next section, but for values of $x$ lower than $x_{AF}$, the behaviour is unaffected by these details.}

 In the finite $N$ cases,  the influence of the single trace coupling $z_1$  cannot be ignored on the question of complete asymptotic freedom, and the analysis of \cite{Pica:2016krb} needs to be expanded  to include multiple quartic self-couplings. An in-depth analysis  goes beyond the scope of this work. Nevertheless, by continuity we expect, at least for $N$ and $p$ sufficiently large, the theory to still feature a complete asymptotically free region in coupling space.

\subsection{Comments on Asymptotic Safety}
We saw in Fig. \ref{fig:PhaseDiagramGG-AS} that hints of asymptotic safety show up in the gauge-Fermion $N=5$ Generalized Georgi-Glashow theory for high values of $p$. From careful analysis of the conditions for asymptotic safety \cite{Litim:2014uca} one expects the presence of a Yukawa coupling between a flavored meson and gauged fermions to help bring about the presence of asymptotic safety, by lowering the needed values of $p$. The simple reason behind this expectation is the fact that Yukawa interactions along the fixed-flow, where the Yukawa beta function vanishes, contribute negatively to the resulting two-loop coefficient of the gauge beta function.  

\begin{figure}[hbt]
  \includegraphics[width=.85\columnwidth]{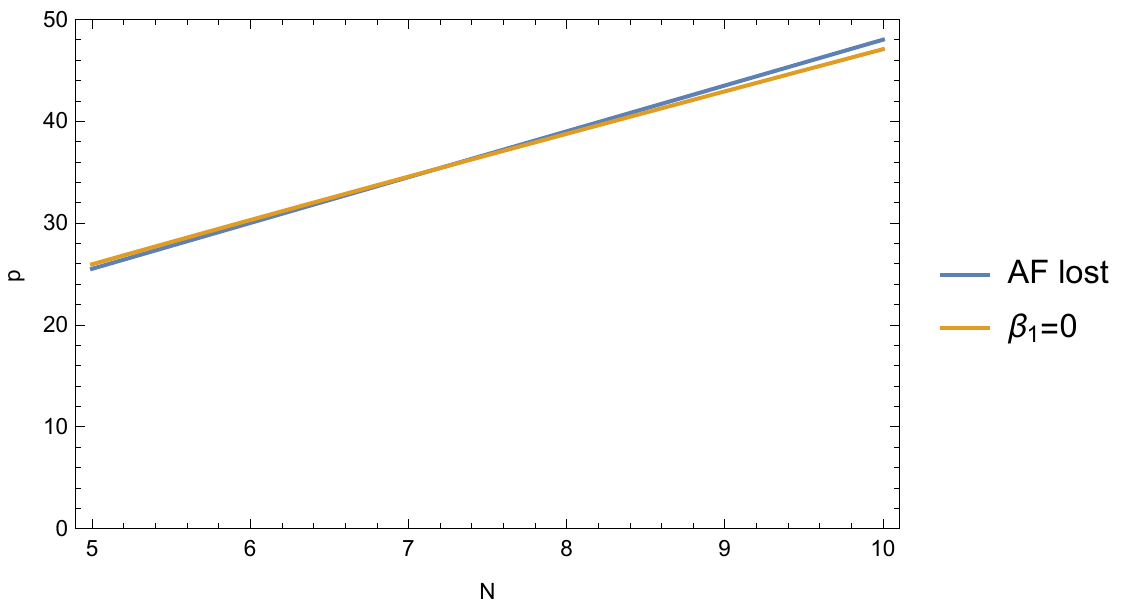}
  \caption{Conformal window for the Generalized Georgi-Glashow theory with mesons and $N=5$ via the two loop order estimate.}\label{fig:PhaseDiagramGGM-pN}
\end{figure}

To elucidate this point here, we write the  $a_M$ beta function to one loop for the generalized GG model with the meson  field $M$ and find the fixed-flow by setting it to zero \ea{
  \beta_{a_M} &= a_M^2(3N+2p-4)-6a_g a_M \frac{N^2-1}{N}\\
  a_M^* &= \frac{6}{3N+2p-4} \frac{N^2-1}{N}a_g
}
the resulting two-loop effective gauge beta function reads:
\ea{
  \beta_{a_g}^{eff} &= \left(-4-6 N+\frac{4 p}{3}\right) a_g^2+\left\{\left(-1-13 N^2-\frac{2 (p-6)}{N}+N \left(-30+\frac{26 p}{3}\right)\right)-\frac{12 p (N+p-4)}{(3 N+2 p-4)}\frac{N^2-1}{N}\right\} a_g^3\\
  &= \left(\frac{4 p}{3}-34\right) a_g^2+\left\{\frac{4}{15} (161 p-1776)-\frac{288 p (1+p)}{5 (11+2 p)}\right\} a_g^3  
}
where in the second equation we have set $N=5$. In Fig \ref{fig:PhaseDiagramGGM-pN}, we plot simultaneously when the first (blue) and the second (orange) coefficient vanish. Since the second coefficient is negative below the orange curve we deduce that for $N=5,6,7$ a conformal window for asymptotic safety opens up albeit for a tiny region of non-integer $p$ for integer $N$. 

\begin{figure}[hbt]
  \subfloat[The fixed point values when the almost-decoupled $z_1$ coupling numerically smallest.]{\includegraphics[width=.45\columnwidth]{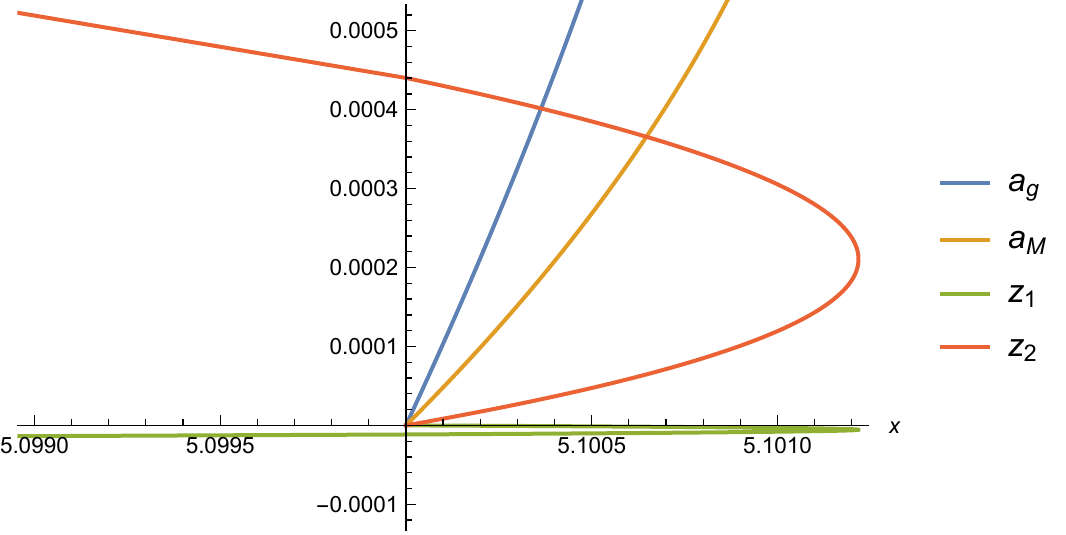}\label{fig:n5ggGY1z}}
  \subfloat[The fixed point values when the almost-decoupled $z_1$ coupling numerically largest.]{\includegraphics[width=.45\columnwidth]{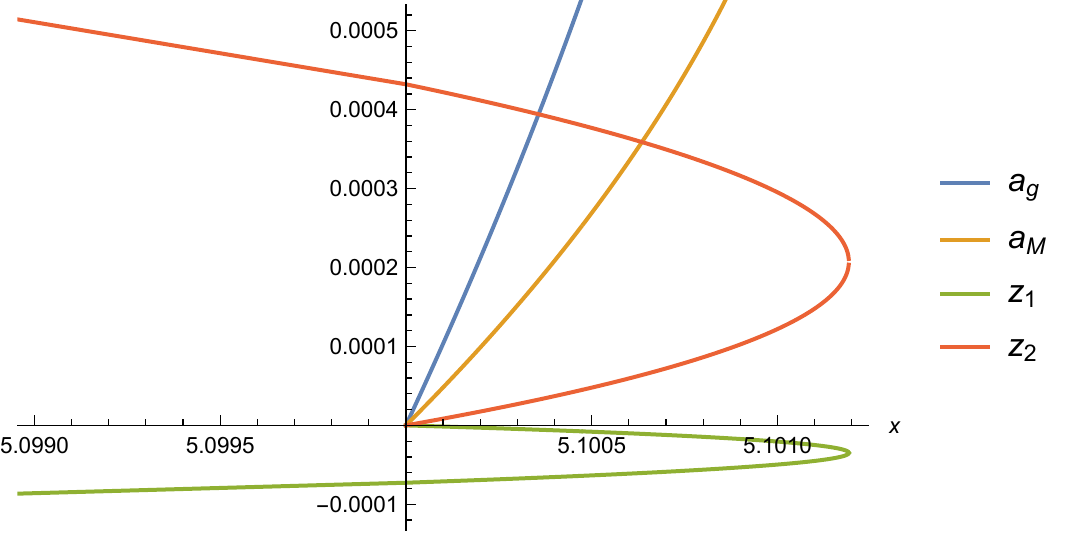}\label{fig:n5ggGY2z}}
  \caption{Fixed points values of the mesonic gauge-Yukawa GG theory with $N=5$.}
\end{figure}

Superficially this seems at odds with the three-loop result found in Figs  \ref{fig:n5ggGY1} and \ref{fig:n5ggGY2} indicating the presence of an IR fixed point for values of $x<x_{AF}=\frac{51}{10}$ corresponding to $p<\frac{51}{2}$. However,  a careful study shows that by zooming into the figures around the point where asymptotic freedom is lost, we find that there is no contradiction ( Figs  \ref{fig:n5ggGY1z} and \ref{fig:n5ggGY2z}). For values of $x$ slightly larger than $x_{AF}$ (corresponding to non-integer $p$), we do have asymptotic safety, but the fixed point soon turns around and yields a perturbative IR-fixed point for $x<x_{AF}$. The turning point is at $x=x^*\approx 5.10122$.
 
The analysis shows how the presence of a scalar degree of freedom, even if singlet under the gauge interactions, greatly changes the phase diagram structure with respect to the pure gauge-Fermion chiral gauge theory.

\section{Chiral gauge theories with a Higgs-like scalar}
In this section, we will include a scalar $H$ transforming according to the fundamental representation of the gauge group \emph{instead} of the mesonic singlet field $M$. This means that the Lagrangian will be extended to include
\ea{
  \mc{L} &= \mc L_{\chi GT} + \mc L_H \label{Lagrangian-higgs}\\
  \mc L_H &= D_\mu (H^\dagger)_aD^\mu H^a + (y_k T^{\{a,b\}} \tilde F^k_a H_b + h.c.) + \lambda (H^\dagger)_aH^a(H^\dagger)_bH^b.
}
Here we adopt the convention that $y_k$ is a vector where the first entry is $y_H$ and all others zero,\footnote{This may seem like a very limiting condition, but it is related to e.g. setting all entries equal the same value by an SU($N\pm4+p$)-transformation.} such that $y_ky^k=y_H^2$. 

We rescale the newly introduced coupling constants in the following manner
\ea{
  a_H = \frac{y_H^2}{(4\pi)^2} \ , \qquad a_\lambda = \frac{\lambda}{(4\pi)^2}.
}

Since we can, in this case, only form a single quartic coupling, the theory has only three beta functions. We work here at finite $N$ and $p$, and list the full beta functions in Appendix \ref{sec:xGT-higgs-betas}.

\begin{figure}[hbt]
  \subfloat[Fixed points values of the higgs-like gauge-Yukawa BY theory with $N=5$. Note the fixed point merger marking the lower boundary of the conformal window at $x\approx1.2$.]{\includegraphics[width=.45\columnwidth]{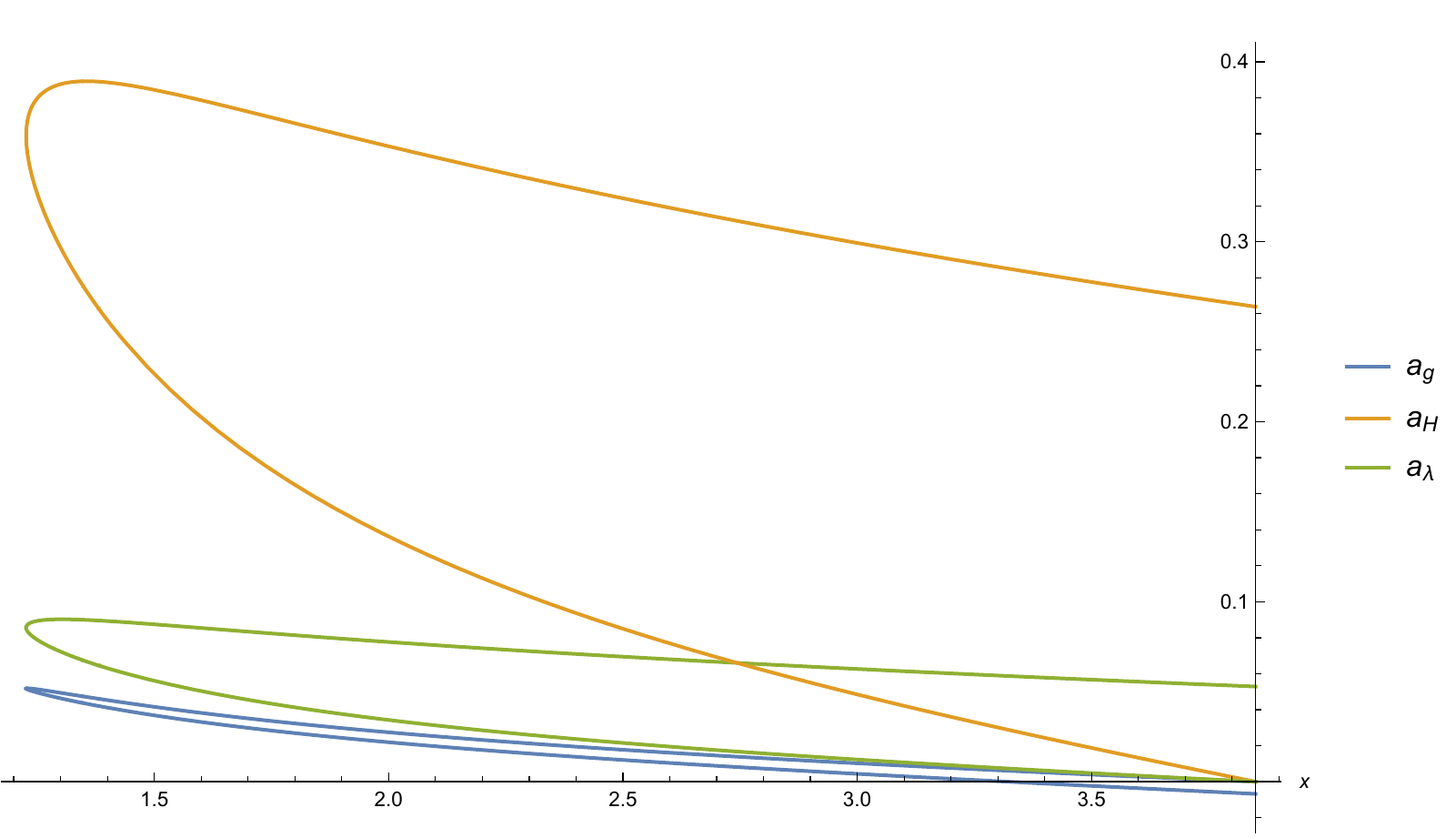}\label{fig:n5byGH-merge}}
  \subfloat[Fixed points values of the higgs-like gauge-Yukawa GG theory with $N=5$. Note the fixed point merger marking the lower boundary of the conformal window at $x\approx3.6$.]{\includegraphics[width=.45\columnwidth]{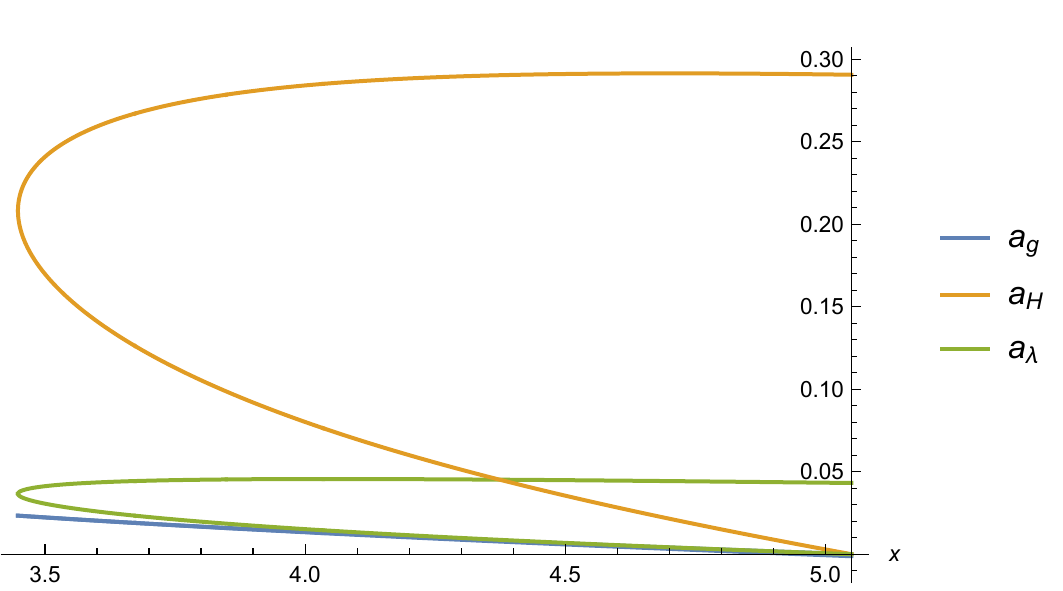}\label{fig:n5ggGH-merge}}
  \caption{The higgs-like chiral gauge-Yukawa theories with $N=5$.}
\end{figure}

We learn that the presence of this specific scalar matter does little to change the basic picture found in the pure gauge-Fermion case (see Section \ref{sec:gaurge-fermion}) at the 2 loop level since the contribution of charged scalar degrees of freedom enters the gauge beta function with the opposite sign of the Yukawa interactions. One notable feature that occurs at the 3 loop level, however, is that we observe a fixed-point merger which provides a calculable lower boundary to the asymptotically free conformal window (see Figures \ref{fig:n5byGH-merge} and \ref{fig:n5ggGH-merge}). Therefore, conformality will be lost smoothly, and we expect that a walking region will be present for $x$ slightly below the merger value. A careful analysis of a similar situation was performed in \cite{Antipin:2012kc} and we will not discuss this phenomenon further in this paper. 

\subsection{Complete Asymptotic Freedom}
\begin{figure}[bt]
  \subfloat[Area of the parameter space of BY theory where complete asymptotic freedom can be found. The shaded region is completely asymptotically free in both the fixed-flow and $a_y\to0$ limits.]{\includegraphics[width=.45\columnwidth]{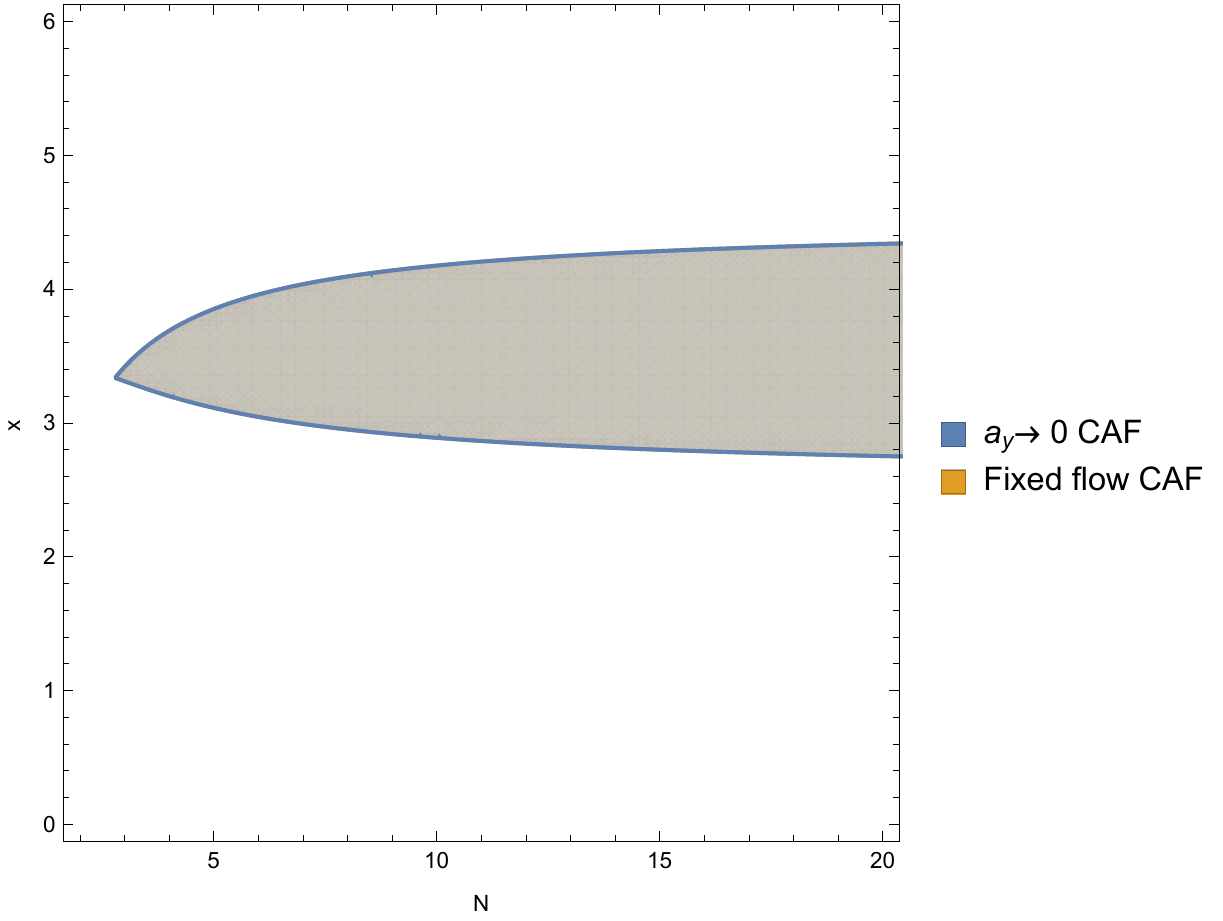}\label{fig:caf-cw-byh5}}
  \subfloat[Area of the parameter space of GG theory where complete asymptotic freedom can be found. The upper shaded shaded region is completely asymptotically free in both the fixed-flow and $a_y\to0$ limits, whereas the lower only is in the fixed-flow limit.]{\includegraphics[width=.45\columnwidth]{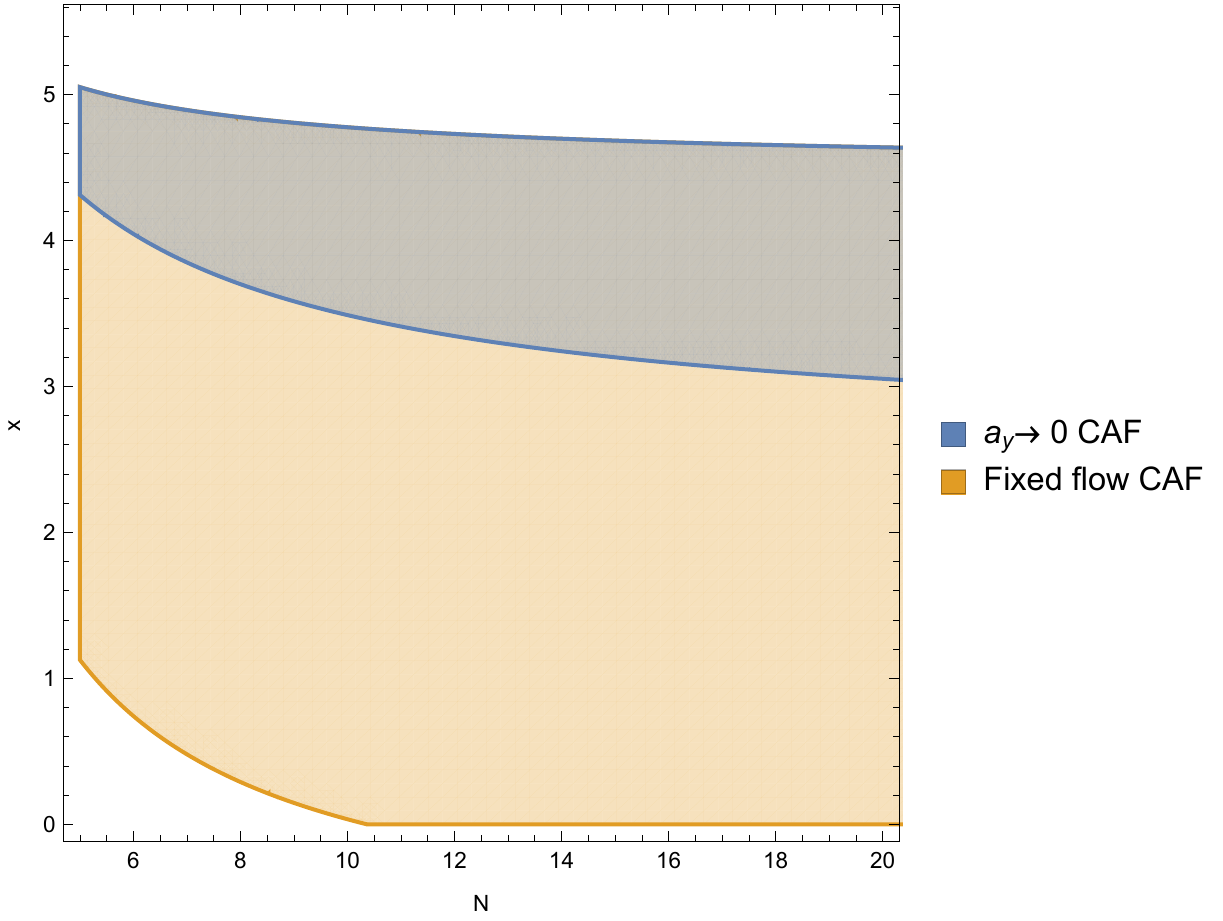}\label{fig:caf-cw-ggh5}}
  \caption{Complete asymptotic freedom in chiral gauge theories with a Higgs-like scalar.}
\end{figure}

Since, by construction, we have one gauge, one Yukawa, and one quartic coupling, we can perform the complete asymptotic freedom (CAF) analysis at any $N$. This is neatly summarised in terms of the CAF parameter space regions of the theory in Figures \ref{fig:caf-cw-byh5} and \ref{fig:caf-cw-ggh5}. 

We see that both the BY and GG theories exhibit complete asymptotic freedom for certain values of $N$ and $p$. In the BY theory (Figure \ref{fig:caf-cw-byh5}), the region of paramter space is smaller, but in the entire region, CAF can be realized in both the $a_y\to0$ and fixed-flow limits, and (presumably) for any value of the yukawa coupling between those two extremes. Conversely, in the GG theory, CAF is realized for a large swath of parameter space, but for most of it, the $a_y\to0$ limit does not allow for complete asymptotic freedom. 

\section{The generalized Georgi-Glashow model with all scalars}
At last we consider the generalized Georgi-Glashow theory featuring simultaneously both the mesonic and higgs-like scalars. 

\ea{
  \mc{L} ={}& \mc L_{\chi GT} + \mc L_H + \mc L_M \label{Lagrangian-higgs-meson}\\
  \mc L_H ={}& D_\mu (H^\dagger)_aD^\mu H^a + (y_H f_k T^{\{a,b\}} \tilde F^k_a H_b + h.c.) + \lambda (H^\dagger)_aH^a(H^\dagger)_bH^b.\\
  \begin{split}
  \mc L_M ={}& \Tr[\p_\mu M^\dagger\p^\mu M]+(y_M(\delta^j_k-f^jf_k) \tilde F_j M^k_l F^l + h.c.) +(y_1f^jf_k \tilde F_j M^k_l F^l + h.c.)\\
  & + u_1 \Tr[M^\dagger M]\Tr[M^\dagger M] + u_2 \Tr[M^\dagger M M^\dagger M],
  \end{split}
}
where we have made some slight changes to the form of the Lagrangian compared to the mesonic and Higgs-like Lagrangian considered previously. Firstly, we have made explicit the fact that $y_k=y_H f_k$ where 
\ea{
  f_k=\begin{cases} 1 & k=1\\0 & k\neq1\end{cases}
}
The Higgs-like Yukawa interaction then breaks the previous symmetry of the mesonic Yukawa coupling into the two pieces shown above through loop corrections. This comes about because only $\tilde F^1$ couples to the Higgs field $H$, but all $\tilde F^k$ couple to the mesonic field $M$.

In analogy with our previous analysis, we rescale the couplings
\ea{
  a_g = \frac{g^2}{(4\pi)^2}&&a_H = \frac{y_H^2}{(4\pi)^2}&&a_M = \frac{y_M^2}{(4\pi)^2}&&a_1 = \frac{y_1^2}{(4\pi)^2},
}
and the beta functions up to two loops in the gauge coupling and one loop in the Yukawas are given by
\ea{
  \begin{split}
  \beta_{a_g} ={}& -a_g^2\left(\frac{11}{3}+6N-\frac{4Nx}{3}\right)-a_g^3\left(-\frac{10}{N}+1+2x+\frac{82N}{3}+13N^2-\frac{26N^2x}{3}\right)\\
  &+a_g^2a_H\left(\frac{5}{2}-\frac{3N}{2}\right)+a_g^2a_M\left(10Nx-2N^2(x+x^2)\right)-2a_g^2a_1 N x.
  \end{split}\\
  \beta_{a_H} ={}& a_ga_H\left(\frac{15}{N}+6-9N\right)-\frac{a_H^2}{2}(1-3N)+a_Ha_1Nx\\
  \beta_{a_M} ={}& a_ga_M\left(\frac{6}{N}-6N\right)-a_M^2\left(5-N\left(3+2x\right)\right)+a_Ma_1\\
  \beta_{a_1} ={}& a_ga_1\left(\frac{6}{N}-6N\right)-a_Ha_1\left(\frac12-\frac N2\right)-a_Ma_1\left(5-N(1+x)\right)+a_1^2\left(1+N\left(2+x\right)\right).
}

\begin{figure}[hbt]
  \centering
  \includegraphics[width=.85\textwidth]{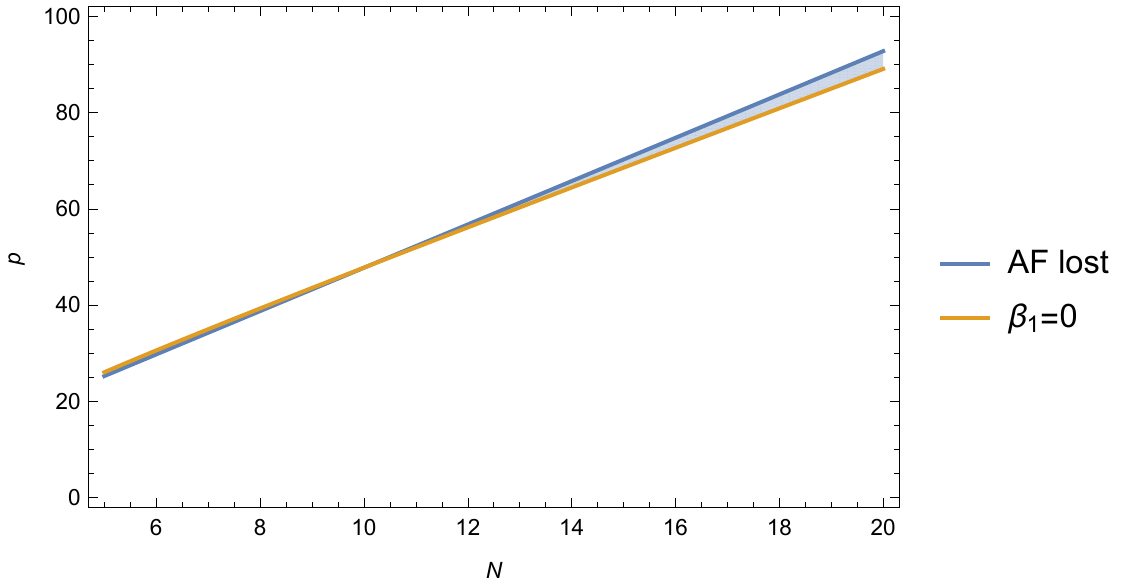}
  \caption{Conformal window of the generalized Georgi-Glashow theory with higgs- and meson-like scalars. The blue line defines the border between asymptotic and infrared freedom, the yellow where $\beta_1=0$. The asymptotically free conformal window is shaded in blue and the asymptotically safe is shaded in yellow.}
  \label{fig:cw-gghm}
\end{figure}

We find the conformal window of this theory using the simplest possible criteria, i.e. that the border of asymptotic freedom determines one edge of the conformal window and the vanishing of the effective two-loop coefficient the second, see Fig. \ref{fig:cw-gghm}. To find the effective two-loop coefficient, we find the fixed point values for $a_H,a_M$ and $a_1$ using $\beta_{a_H}=\beta_{a_M}=\beta_{a_1}=0$.

We observe that there the theory seems to exhibit two qualitatively different conformal windows. For $N>10$, there is a narrow, but widening as $N$ increases, slice of parameter space where conformality can be found within the asymptotically free region of parameter space. For $N\leq10$, however, we find that the conformal window lies above the boundary of asymptotic freedom, meaning that any fixed points will be asymptotically safe. Careful examination shows that asymptotically safe fixed points exist for four distinct theories given by ($N=5,p=26$), ($N=6,p=30$), and ($N=8,p=39$). The asymptotically safe conformal window also extends to $N=7$, $N=9$ and $N=10$, however here there are no integer values of $p$ for which asymptotic safety can be realised.

\section{Concluding remarks}

We studied the phase diagram of relevant chiral gauge-Yukawa theories in perturbation theory with and without several scalar degrees of freedom transforming according to distinct representations of the underlying gauge group. The gauge-fermion sector corresponds to the generalized Georgi-Glashow and Bars-Yankielowicz theories. Not only did we unveil the phase diagram of these theories, but we further disentangled their ultraviolet asymptotic nature according to whether they are asymptotically free or safe.  

The emerging general picture is that it is possible to have complete asymptotically free chiral gauge field theories with scalars  and further, that these theories can have a controllable infrared conformal window.

Asymptotic safety can kick in, once asymptotic freedom is lost in the gauge coupling, non-perturbatively when scalars are absent and furthermore above a critical number of flavours in agreement with the observations made in \cite{SanninoERG2016}.  When, however, scalar singlets are present Yukawa interactions help taming the ultraviolet behaviour of the gauge interactions and perturbative asymptotic safety emerges as observed first in  \cite{Litim:2014uca}. 
  
  This is well in line with the argument of \cite{Bond:2016dvk} that asymptotic safety can only occur in theories with gauge and Yukawa couplings.

\section*{Acknowledgements}
We would like to thank Elena Vigiani and Giulio Maria Pelaggi for invaluable assistance in double-checking our calculations of the beta functions. The CP$^3$-Origins centre is partially funded by the Danish National Research Foundation, grant number DNRF90.

\appendix
\section{Beta functions and anomalous dimensions}
\subsection{Gauge-fermion theories}
In this appendix, we present the beta functions of the gauge-fermion theories under consideration. The beta functions are derived on the basis of References \cite{Machacek:1983tz, Machacek:1983fi, Machacek:1984zw,Luo:2002ti,Pickering:2001aq,Molgaard:2014hpa}, which is done in the Landau gauge of the $\overline{\mathrm{MS}}$ scheme and is as such independent of the gauge-fixing parameter. However, if one considers the theory in another scheme, more care must be taken to ensure gauge invariance, see e.g. \cite{Ryttov:2012nt,Ryttov:2016hal}.

To make our expressions more transparent, we will work initially with the coupling
\ea{
  a_g = \frac{g^2}{(4\pi)^2}. 
}

To the three-loop order, the beta function in generalized Bars-Yankielowicz and Georgi-Glashow theory is:
\ea{
\begin{split}
  \beta_{a_g} ={}& -a_g^2 \left\{\left(6-\frac{4 x}{3}\right) N\mp 4\right\}-a_g^3 \left\{\left(13-\frac{26 x}{3}\right) N^2\mp 30 N+(1+2 x)\pm \frac{12}{N}\right\}\\
  &-a_g^4 \left\{\left(\frac{127}{3}-\frac{979 x}{18}+\frac{112 x^2}{27}\right) N^3\mp \left(180-\frac{82 x}{3}\right) N^2+\left(\frac{201}{4}+\frac{77 x}{9}-\frac{11 x^2}{9}\right) N\right.\\
 &\left.\pm \left(\frac{283}{6}-11 x\right)-\frac{103-2 x}{4 N}\pm
   \frac{9}{N^2}\right\},
\end{split} \label{eq:by-gg-fermion-beta}
}
where the upper (lower) signs correspond to the generalized BY (GG) theory. $N$ is the number of colors which is restricted to $N>5$ for GG theory, $x=p/N$ is a more convenient variable than $p$ when considering the large N limit and it is a simple matter to make the replacement if one cares only about a specific finite $N$.

We can also compute the anomalous dimension of the fermion mass operator $F\tilde F$ to two loop order
\ea{
  \gamma_{F\tilde F} &= a_g\left\{\frac{3 N}{2}-\frac{3}{2 N}\right\}+a_g^2\left\{\left(\frac{61}{8}-\frac{5 x}{6}\right) N^2\mp \frac{15 N}{6}+\left(-8+\frac{5 x}{6}\right)\pm \frac{15}{6 N}+\frac{3}{8 N^2}\right\}
}
where the upper (lower) signs again correspond to the generalized BY (GG) theory.

%

\subsection{Chiral gauge theories with a mesonic-like scalar}\label{sec:xGT-mes-betas}
The following are the beta functions for the chiral gauge theories (either BY or GG) that include a mesonic-scalar like operator with the Lagrangian given in \eqref{Lagrangian-meson}. 
\ea{
\begin{split}
  \beta_{a_g}={}&- a_g^2 \left\{\left(6-\frac{4 x}{3}\right) N\mp 4\right\}-a_g^3 \left\{\left(13-\frac{26 x}{3}\right) N^2\mp 30 N+(1+2 x)\pm \frac{12}{N}\right\}-a_g^2 a_M \Big\{2 x (1+x) N^2\\
  &\pm 8 x N\Big\}-a_g^4 \left\{\left(\frac{127}{3}-\frac{979 x}{18}+\frac{112 x^2}{27}\right) N^3\mp
   \left(180-\frac{82 x}{3}\right) N^2+\left(\frac{201}{4}+\frac{77 x}{9}-\frac{11 x^2}{9}\right) N\right.\\
  &\left.\pm \left(\frac{283}{6}-11x\right)+\left(\frac{-103+2 x}{4 N}\pm \frac{9}{N^2}\right)\right\}-a_g^3 a_M \left\{\frac{27 x}{2} (1+x) N^3\pm 54 x N^2-\frac{3 x}{2}(1+x) N\mp 6 x\right\}\\
  &-a_g^2 a_M^2 \left\{-x \left(5+8 x+3 x^2\right) N^3\mp 2 x (13+9 x) N^2-24 N x\right\}
\end{split}\label{eq:xGT-beta-ag}\\
\begin{split}
  \beta_{a_M}={}& a_g a_M \left\{-6 N+\frac{6}{N}\right\}+a_M^2\{(3+2 x) N\pm 4\}+a_g^2 a_M \bigg\{\left(-\frac{61}{2}+\frac{10 x}{3}\right) N^2\pm 10 N+\left(32-\frac{10x}{3}\right)\mp \frac{10}{N}\\
  &-\frac{3}{2 N^2}\bigg\}+a_g a_M^2 \left\{(9+8 x) N^2\pm 16 N-(9+8 x)\mp \frac{16}{N}\right\}+a_M^3\Bigg\{-\left(3+\frac{13 x}{2}+\frac{x^2}{2}\right) N^2\\
  &\mp(12+2 x)N+4\Bigg\}+a_M^2 z_1 \{-8 (1+2 x) N\mp 32\{+a_M^2 z_2 \left\{-8 x (1+x) N^2\mp 32 x
   N-8\right\}\\
  &+a_M z_1^2 \left\{4 x (1+x) N^2\pm 16 x N+4\right\}+a_M z_1 z_2\{8 (1+2 x) N\pm 32\}\\
  &+a_M z_2^2 \left\{4 x (1+x) N^2\pm 16 x N+4\right\}
\end{split}\label{eq:xGT-beta-aM}\\
  \beta_{z_1} ={}& 4 N a_M z_1+z_1^2 \left\{4 x (1+x) N^2\pm 16 x N+16\right\}+z_1 z_2\{8 (1+2 x) N\pm 32\}+12 z_2^2\label{eq:xGT-beta-z1}\\
  \beta_{z_2} ={}& -2 N a_M^2+4 N a_M z_2+24 z_1 z_2+ z_2^2\{4 (1+2 x) N\pm 16\}\label{eq:xGT-beta-z2}
}

\subsection{Chiral gauge theories with a higgs-like scalar}\label{sec:xGT-higgs-betas}
The following are the beta functions for the chiral gauge theories (either BY or GG) that include a higgs-like scalar operator with the Lagrangian given in \eqref{Lagrangian-higgs}. 
\ea{
\begin{split}
  \beta_{a_g}={}& a_g^2 \left\{\left(\frac{4 x}{3}-6\right) N+\frac{1\pm 12}{3}\right\}+a_g^3 \left\{\left(\frac{26 x}{3}-13\right) N^2+\frac{8\pm 90}{3}N-(1+2 x)-\frac{2\pm 12}{N}\right\}\\
  &-a_g^2 a_H \left\{\frac{3 N}{2}\pm \frac{5}{2}\right\}+a_g^4 \left\{\left(-\frac{127}{3}+\frac{979
   x}{18}-\frac{112 x^2}{27}\right) N^3+\left(\frac{1507\pm 12960}{72}\right.\right.\\
  &\left.\left.-\frac{335\pm 2952}{108}x\right) N^2+\left(-\frac{10997\pm 2286}{216}-\frac{77 x}{9}+\frac{11 x^2}{9}\right) N-\frac{382\pm 849}{18}+\frac{73\pm 396}{36}x\right.\\
  &\left.+\frac{1903\pm 576-36x}{72 N}+\frac{29\mp 72}{8 N^2}\right\}+a_g^3 a_H \left\{-\frac{261 N^2}{16}\mp \frac{465N}{16}+\frac{133}{16}\pm \frac{281}{16 N}\right\}\\
  &+a_g^3 a_{\lambda } \left\{N+2-\frac{2}{N}\right\}+a_g^2 a_H^2\left\{\frac{57 N^2}{32}+\frac{13\pm 82}{16}N+\frac{115\pm 38}{32}\right\}+a_g^2 a_{\lambda }^2 \{-2 N-2\}
\end{split}\\
\begin{split}
  \beta_{a_H}={}& a_g a_H \left\{-9 N\mp 6+\frac{15}{N}\right\}+a_H^2 \left\{\frac{3 N}{2}+\frac{2\pm3}{2}\right\}+a_g^2 a_H \left\{\left(-\frac{129}{4}+4x\right) N^2+\left(\frac{5\mp 34}{2}\pm\frac{10 x}{3}\right) N\right.\\
  &\left.+\frac{819\pm 22}{12}-\frac{22 x}{3}-\frac{13\pm 48}{3 N}-\frac{3}{N^2}\right\}+a_g a_H^2 \left\{\frac{63}{8}N^2+\frac{30\mp153}{8}N-\frac{39\pm4}{8}-\frac{26\pm153}{8N}\right\}\\
  &+a_H^3 \left\{-\frac{3N^2}{4}-\frac{7\pm2}{4}N+\frac{9\mp7}{4}\right\}+a_H^2 a_{\lambda } \{-4N-4(2\pm 1)\}+a_H a_{\lambda }^2 \{4+4N\}
\end{split}\\
\begin{split}
  \beta_{a_\lambda}={}& a_g^2 \left\{\frac{3 N}{4}+\frac{3}{4}-\frac{3}{N}+\frac{3}{2 N^2}\right\}+a_g a_{\lambda } \left\{-6 N+\frac{6}{N}\right\}+a_H^2
   \left\{-\frac{N}{2}-\frac{1\pm 2}{2}\right\}+a_H a_{\lambda } \left\{2N\pm2\right\}\\
  &+a_{\lambda }^2 \{4 N+16\}
\end{split}
}

%

\section{Summary of Complete Asymptotic Freedom Conditions}\label{sec:summary-caf}
 The CAF conditions can be identified at one loop in all couplings. The gauge coupling evolution at one loop reads
\begin{eqnarray}\label{gauge}
\mu \frac{d \alpha_g}{d \mu} &=& b_0 \alpha_g^2 \ .
\end{eqnarray}
For a single Yukawa coupling is 
\begin{eqnarray}\label{Yukawa}
\mu \frac{d \alpha_H}{d \mu} &=&  \alpha_H \left[ c_1 \alpha_g + c_2 \alpha_H \right] 
\end{eqnarray}
where in general $c_1<0$ and $c_2>0$ while the scalar self-coupling reads 
\begin{eqnarray}
\mu \frac{d\alpha_{\lambda}}{d\mu} &=& \alpha_{\lambda} \left( d_1\alpha_{\lambda} + d_2 \alpha_g + d_3 \alpha_H  \right) + d_4 \alpha_g^2 + d_5 \alpha_H^2
\end{eqnarray}
where $d_1,d_3,d_4\geq 0$ and $d_2,d_5\leq 0$. Together with Eq. \ref{gauge} and \ref{Yukawa} it describes the running of the gauge, Yukawa and self coupling in a general gauge-Yukawa system at one loop order.  
 
If the gauge and Yukawa couplings are not on their fixed flow these conditions are
\begin{eqnarray}
b_0 <0  \ , \qquad b_0-c_1 >0 \ , \qquad k \geq 0 \ , \qquad b_0 - d_2 +\sqrt{k}>0 \ , \qquad \text{Condition CAF}_1
\end{eqnarray}
where 
\ea{
  k = \left(b_0-d_2\right)^2 -4d_1d_4
}
If the beta function coefficients satisfy these constraints and the couplings satisfy appropriate initial (infrared) conditions the theory is \emph{complete asymptotically free}. The first (second) condition is necessary to ensure asymptotic freedom of the gauge (Yukawa) coupling while the third and fourth conditions are necessary to ensure asymptotic freedom and positivity of the self coupling. 

On the other hand if the gauge and Yukawa couplings are on their fixed flow then the necessary set of conditions that the beta function coefficients must satisfy is
\begin{eqnarray}
b_0 <0  \ , \qquad b_0-c_1 >0 \ , \qquad k' \geq 0 \ , \qquad b_0 - d_2' + \sqrt{k'}>0 \ , \qquad \text{Condition CAF}_2
\end{eqnarray}
where
\ea{
  d_2' &= d_2+d_3\frac{b_0-c_1}{c_2}\\
  k' &= \left(b_0-d_2-d_3\frac{b_0-c_1}{c_2}\right)^2 -4d_1\left(d_4+d_5\left(\frac{b_0-c_1}{c_2}\right)^2\right)
}
The condition for asymptotic freedom of the self coupling is in this case different from the condition where the gauge and Yukawa couplings are not on their fixed flow. This is because the running of the Yukawa coupling can no  longer be neglected and has an influence on the running of the self coupling. If these contions CAF$_2$ are satisfied and the couplings satisfy appropriate initial (infrared) conditions the theory is \emph{complete asymptotically free}. 

Investigations of asymptotically free scenarios  in non-abelian Higgs  models making use of nonperturbative approaches appeared in \cite{Gies:2015lia,Gies:2016kkk}.


\end{document}